\documentclass[sigconf]{acmart}%
\usepackage{color}%
\newcommand{\todo}[1]{{\leavevmode\color{black}#1}}%
\newcommand{\ok}[1]{{\leavevmode\color{black}#1}}%
\newcommand{\transition}[1]{{\leavevmode\color{black}#1}}%
\newcommand{\rev}[1]{{\leavevmode\color{black}#1}}%
\usepackage{multirow}%
\usepackage[normalem]{ulem}%
\usepackage{tabularx}%
\usepackage{caption}%
\usepackage{subcaption}%
\usepackage[strings]{underscore}%
\usepackage[section]{placeins}%
\usepackage{multicol}%
\usepackage{etoolbox}%
\usepackage{ragged2e}%
\newcolumntype{Y}{>{\RaggedRight\arraybackslash}X}%
\usepackage[utf8]{inputenc} 
\copyrightyear{2026}
\acmYear{2026}
\setcopyright{cc}
\setcctype{by}
\acmConference[CHI '26]{Proceedings of the 2026 CHI Conference on Human Factors in Computing Systems}{April 13--17, 2026}{Barcelona, Spain}
\acmBooktitle{Proceedings of the 2026 CHI Conference on Human Factors in Computing Systems (CHI '26), April 13--17, 2026, Barcelona, Spain}
\acmPrice{}
\acmDOI{10.1145/3772318.3790681}
\acmISBN{979-8-4007-2278-3/2026/04}
\begin{document}%
\title[Default Images]{%
An Exploration of Default Images in Text-to-Image Generation
}%


\author{Hannu Simonen}%
\email{hannu.simonen@student.oulu.fi}%
\affiliation{%
  \institution{University of Oulu}
  \city{Oulu}
  \country{Finland}
}%

\author{Atte Kiviniemi}%
\email{atte.kiviniemi@student.oulu.fi}%
\affiliation{%
  \institution{University of Oulu}
  \city{Oulu}
  \country{Finland}
}%

\author{Hannah Johnston}%
\email{hannahjohnston@cmail.carleton.ca}%
\orcid{0009-0001-9599-6546}
\affiliation{%
  \institution{Carleton University}
  \city{Ottawa}
  \country{Canada}
}%

\author{Helena Barranha}
\email{helenabarranha@tecnico.ulisboa.pt}
\orcid{0000-0003-0250-1020}
\affiliation{%
  \institution{IST, University of Lisbon \& 
    IHA-NOVA FCSH / IN2PAST
  }
  \city{Lisbon}
  \country{Portugal}
}

\author{Jonas Oppenlaender}%
\email{jonas.oppenlaender@oulu.fi}%
\orcid{0000-0002-2342-1540}%
\affiliation{%
  \institution{University of Oulu}
  \city{Oulu}
  \country{Finland}
}%


\begin{abstract}
In the creative practice of text-to-image (TTI) generation, images are synthesized from textual prompts. By design, TTI models always yield an output, even if the prompt contains unknown terms. In this case, the model may generate default images: images that closely resemble each other across many unrelated prompts. Studying default images is valuable for designing better solutions for prompt engineering and TTI generation. We present the first investigation into default images on Midjourney. We describe an initial study in which we manually created input prompts triggering default images, and several ablation studies. Building on these, we conduct a computational analysis of \rev{over 750,000 images}, revealing consistent default images across unrelated prompts. We also conduct an online user study investigating how default images may affect user satisfaction. Our work lays the foundation for understanding default images in TTI generation, highlighting their practical relevance as well as challenges and future research directions.
\end{abstract}%
%
\begin{teaserfigure}
\centering
    \includegraphics[width=\textwidth]{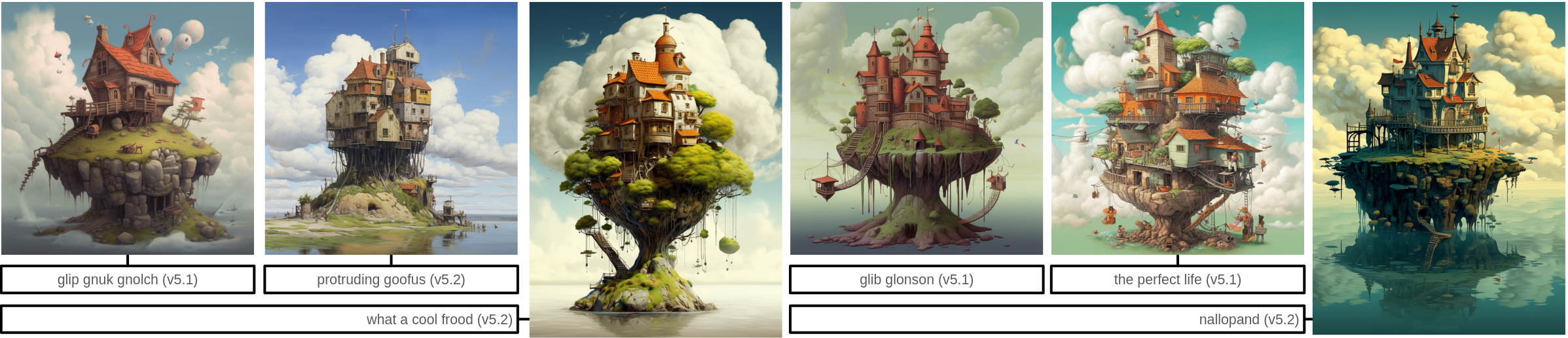}%
\\
    \includegraphics[width=\textwidth]{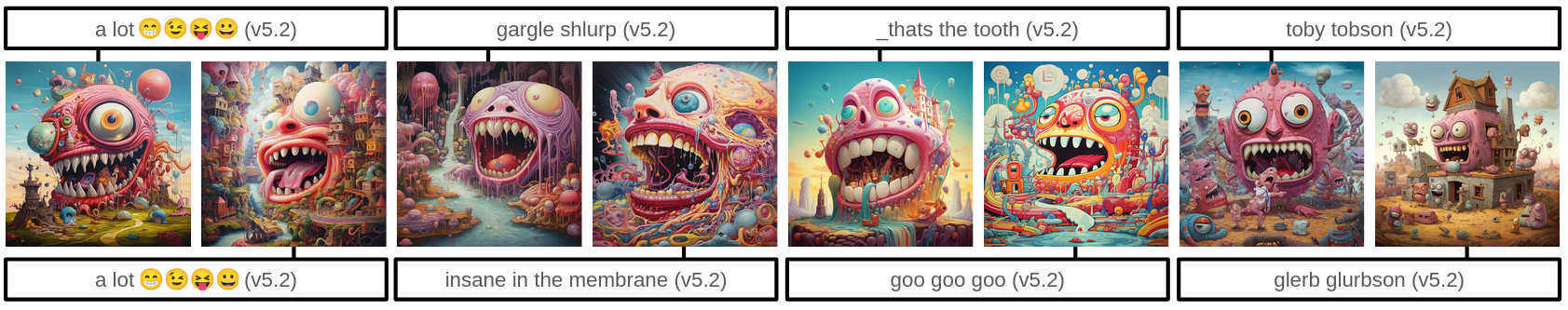}%
    \caption{
    \rev{Default images in Midjourney.
    Varied prompts lead to visually similar outputs, motivating our exploration of default image behavior.
    }%
    }
    \Description{The figure shows different default images, resulting from unrelated input prompts. It is clearly visible that these images have a common theme, motif (i.e., subjects), and style.}
    \label{fig:teaser}
\end{teaserfigure}
\maketitle
%
%
\section{Introduction}%
Generative artificial intelligence (AI) has swept into mainstream.
One 
popular type of generative AI is text-to-image (TTI) generation, where images are created by prompting the generative model with text~\cite{prompting-ai-art,palgrave,feng-etal-2023-uncovering}.
The prompt consists of a textual description of the envisioned image, including keywords that are known to trigger specific subjects, forms, and styles~\cite{taxonomy,ghibli,3491102.3501825.pdf}.
The practice of `prompt engineering' is key to TTI generation~\cite{palgrave}.
Practitioners craft a prompt, observe the outcome, and adjust the prompt until a satisfactory outcome is achieved~\cite{prompting-ai-art}.

While there is a growing number of studies investigating
the practice of prompt engineering or prompt design for TTI generation (e.g., \cite{prompting-ai-art,3613904.3642861.pdf,3491102.3501825.pdf,3527927.3532792.pdf,3706598.3713166.pdf,ICCC-2023_paper_104.pdf,3613905.3650947.pdf}),
and the design of user interfaces for prompt engineering (e.g., \cite{Promptify,3613904.3642016.pdf,2508.06065.pdf,GenAssist}),
    less attention has been given to how models behave when prompts are misunderstood or ambiguously interpreted.
TTI models are trained to always yield an output, even if the input prompt contains terms and concepts not known to the model.
When such a  prompt is given to the TTI model, it may generate what we call \textit{default images} --
    visually similar images resulting from dissimilar prompts (see figures \ref{fig:teaser} and \ref{fig:examples-variations},
and Appendix \ref{fig:more-results}).
These images are seemingly unrelated to the input prompt and, therefore, often do not match the user's expectations.



Default images reveal gaps in the generative model's vocabulary and 
capabilities. 
    For instance, a creative practitioner may prompt a TTI model with the Tagalog word \textit{``pagpag''} (referring to a dish made from leftover food scavenged from garbage dumps), resulting in a default image (e.g., a painting of a woman with a cloud covering her face).
        This indicates that the prompt was not understood and the practitioner should
        revise the wording to improve the generated image.
    Default images can 
        put an early end to a potentially unfruitful pathway in the practitioner's prompting of the TTI model, but they can also
    derail the entire creative process, leading the user to abandon the quest to generate their desired image (cf. Section~\ref{sec:inportance}).
Therefore, default images and their recognition play a crucial role in the creative practice of prompt engineering 
and for discovering the limits of TTI models and their (computational) creativity.
Understanding how generative systems respond to unfamiliar or ill-formed prompts is essential for designing interactive systems that support creative exploration and for providing feedback to users.
In human–AI co-creation, unexpected outputs (such as default images) can disrupt the feedback loop between user intent and system output, introducing ambiguity or undifferentiation. 
This undermines trust, \rev{perceived accuracy, usability,} and user agency \cite{Kreminski2025,Honeycutt202063,Riveiro2021}.
The misalignment between user intent and outputs can lead to frustration and a sense of lost control over the creative process \cite{Louie2020}.
Investigating the phenomenon of default images contributes to broader HCI discussions on transparency, interpretability, and the design of meaningful interactions with generative models.
Default images have 
not received 
attention in prior research.

In this paper, we present the first 
exploration of default images in Midjourney,\footnote{\href{https://www.midjourney.com/}{www.midjourney.com}} a popular image generation system.
We describe our systematic approach to manually create 
    input prompts likely triggering default images and our initial experiments in generating images from these prompts.
We then apply affinity diagramming to find similarities between the generated images and identify 
ten default images.
We further conduct an online user study investigating how default images impact user satisfaction,
and how they align with human perception and expectations.
%
%
Scaling up our analysis, we then present 
insights from a large-scale computational analysis of
    images on Midjourney.

We make
the following contributions:%
\begin{itemize}%
    \item%
    \textit{Conceptual and empirical contribution:}
    We define the concept of \textit{default images} and provide examples of this phenomenon.
    Based on an initial study,
    we contribute six categories of prompts that are likely to elicit default images. We present the results of our manual experiments with these prompts and several ablation studies.
    We also present 
    a user study with crowdsourced participants (N~=~48) investigating how default images may affect user satisfaction.%
    \item
    \textit{Methodological contribution:}
    We describe our initial systematic
        approach for manually creating prompts that trigger default images.
    We then describe a simple method for identifying default images at scale, first clustering images by visual similarity and then filtering by intra-cluster prompt diversity.
    With this method, we computationally analyze 
    \rev{over 750,000}
    images from Midjourney, 
    providing empirical evidence of the existence of default images in response to user-written prompts on Midjourney.
    \item 
    \textit{Open-source contribution:}
    We share 
    a dataset
    with \rev{189,432} prompts 
    and \rev{757,728} candidates of default images,
    enabling future research on default images
    (see \href{https://huggingface.co/datasets/tti-dev/default-images}{https://huggingface.co/ datasets/tti-dev/default-images}).%
\end{itemize}%


We discuss the challenges of identifying unique but realistic input prompts that reliably trigger default images,
and the challenges of identifying default images in a real-world setting.
    Based on the results of our investigation, we make eight postulates, 
    detailing the characteristics of default images in generative TTI models and providing
    a basis for future experimentation and theoretical development.%

%
%
%
%
%
%
%
%
%
\section{Default Images in Text-to-Image Generation}%
%
\subsection{What are Default Images?}%
Some inputs to TTI generation systems may seemingly fail to generate the intended results from the creative practitioner's perspective.
We call this failure mode \textit{default images}. 
We use the term `default' to denote that TTI models resort to generating these images when they do not recognize their input.
This could be because the input terms were not part of the generative model's training dataset or because the terms in the prompt do not invoke strong imagery.
In these cases, distinct prompts may yield near-identical images.
In machine learning, a related concept is called mode collapse \cite{NIPS-2014-generative-adversarial-nets-Paper.pdf}, a failure that limits the usefulness of generative models for image synthesis.

The defining characteristic of default images is that they are encountered in response to multiple different and 
unrelated prompts.
    We demonstrate this in \autoref{fig:teaser} and \autoref{fig:examples-variations}. 
%
While there are often slight variations to the default images, they can be remarkably stable in terms of motif (i.e., subject) and style (cf. \autoref{fig:examples-variations}),
making default images distinctly recognizable.
Experienced practitioners and in particular creative practitioners who seek to deviate from common prompts will likely have encountered default images in their prompting.
This paradoxical situation, which may go unnoticed by most users, reflects the current limitations of TTI models. As \citeauthor{MANOVICH} observes, AI image generators ``deal well with frequently occurring items and patterns in the data but do not know what to do with the infrequent and uncommon'' \cite{MANOVICH}. These models ``are much better at generating visuals of popular subjects and aesthetics than unfamiliar or rare ones.''

Artists and other creative practitioners often seek to generate images that are novel \cite{ko2023large}, a key component of creativity \cite{Runco01012012}.
Default images may pose a particular challenge to these users, since default images with their generic visuals tend to occur with uncommon and highly original prompts.
Default images 
may satisfy users with less specific image requirements or lower task persistence.
In these cases, users may engage in cognitive gap-filling processes~\cite{wertheimer1923gestalt} or construct personal associations~\cite{bartlett1932remembering}, resulting in images that appear more relevant to the prompts than they actually are.
This effect may be further amplified when the generated images have strong aesthetic appeal, since aesthetically pleasing stimuli are often judged more favorably \cite{reber2004processing} and attractive designs are often perceived as more effective, regardless of their actual function \cite{kurosu1995apparent}.
Users may accept visually appealing default outputs rather than investing effort to generate better alternatives.
It is detrimental to creativity overall if TTI users are pushed towards repetitive,
generic images in response to their novel prompting.
This issue makes the phenomenon of default images particularly relevant to visual creativity and originality.

Default images primarily result from how text-to-image models structure internal representations of visual concepts.
When presented with ambiguous or unknown inputs, these models rely on their learned latent space -- the continuous, high-dimensional internal space encoding visual relationships. Regions of this space correspond to generic or frequently occurring visual patterns, which the model defaults to in the absence of specific semantic guidance.
Stochastic sampling adds variability, and over-training or memorization of frequent training examples can reinforce these biases. Ultimately, default images arise from latent space's inherent structure and learned biases.

We can use a metaphor to help understand default images: the latent space as a vast landscape.
Images are generated by sampling from regions within this landscape. Known concepts are easy to spot (think of them as mountains and hills that attract attention). But there are also planes that connect vast regions of the space, which do not correspond strongly to specific learned concepts. Default images result when the TTI model samples from these planes because of the absence of strong `attractors' (known concepts) in the sampled region of latent space.%
%
%
%
%
\subsection{Importance of Studying Default Images}
\label{sec:inportance}
Many Midjourney users type long, descriptive prompts containing specific keywords, representing known concepts and style modifiers \cite{taxonomy,prompting-ai-art}.
Known concepts reliably elicit the desired imagery, but they also act like `magnets' that overshadow any unknown concepts, moving the output away from default images towards imagery related to the known concept. For instance, including \textit{``trump''} or even \textit{``donald''} in a Midjourney prompt may invoke images of a certain U.S. president in many different contexts.
This means the majority of Midjourney users will likely not encounter default images in their image creation efforts.
So why study default images?

Default images provide valuable insights into the robustness and limitations of TTI models.
When models generate similar outputs from diverse, unrelated prompts, this indicates semantic gaps and challenges with linguistic constructs, niche topics, or abstract concepts, and a lack of generalization (i.e., the ability of a model to perform well on unseen data).
Such patterns suggest that the model may rely on memorized outputs rather than generating novel imagery, 
revealing limits in the model's computational creativity.
Awareness of these issues can help with assessing model capabilities, benchmarking TTI model performance, and detecting vulnerabilities related to adversarial prompting.

Understanding default images informs data and model quality improvements.
Default images can indicate that the training data lacks diversity or contains biased patterns, for instance, highlighting insufficient coverage of low-resource languages, such as Tagalog, an Austronesian language spoken as a first language by about a quarter of the population of the Philippines.
Addressing these cases can enhance dataset curation and fine-tuning, promoting more inclusive and culturally diverse 
images.
Studying this phenomenon can help monitor the risk of default images entering training datasets, which could amplify the self-reinforcing loop of AI-generated media \cite{ARIELLI,2505.08803.pdf,2407.17493.pdf}.

Understanding default images is valuable for creative practitioners and prompt engineers exploring computational creativity and comparing different generative models.
Default images reveal the model's limitations in interpreting abstract or unfamiliar concepts. 
Knowledge of these gaps allows users to refine prompts and circumvent ``dead zones'' where outputs are generic or repetitive, and implement strategies to recover from failures, ultimately supporting more meaningful and original visual outputs.

Default images also have potential security implications.
An attacker could infer from the appearance of default images the model's limitations or gaps in training data.
This knowledge could be exploited to reverse-engineer aspects of the model's dataset or architecture.
Default images can act as a model signature, allowing attackers to craft prompts that identify the TTI model or version being used. This information could be help launch more targeted attacks.
For instance, knowing that a model produces a predictable default image for certain prompts could allow attackers to inject misleading or confusing input in systems that rely on generative models.

From a user experience perspective, recognizing patterns in default image generation can inform more resilient and user-friendly interface design.
When users encounter similar outputs from diverse prompts, it can be frustrating. Interfaces that provide feedback, suggest alternative prompts, or guide users through effective prompt engineering can help navigate these limitations and recover from failures, enhancing creativity and usability.%
\section{Related Work}%
%
Our review of related work covers three intertwined topics: 1) how TTI models are built and prompted, 2) known failure modes and ``degenerate'' outputs in generative systems, and 3) user-facing evaluations of generation quality and satisfaction.
%
\subsection{Text-to-Image Generation and Prompt Engineering}%
Text-to-image generation has emerged as a powerful subfield of generative AI, in which deep neural networks (e.g., 
text-conditioned diffusion models) learn to translate free-form textual prompts consisting of discrete words into high-fidelity images. Prominent TTI systems, such as Midjourney, OpenAI’s DALL·E, and Stable Diffusion, demonstrate the ability to produce photorealistic or artistically stylized images across many domains. 
    Generative TTI models are pre-trained on extensive datasets and generate images for a given prompt by probabilistic sampling from their latent space.

While TTI systems attempt to match human expectations, the prompting language is loose and it is near impossible to describe every detail of an envisioned output in one textual prompt, with a limited number of characters \cite{Gafni:2022}.
Further, the model may lack relevant prior knowledge and context.
While special prompt keywords can address some of these shortcomings, such as improving the resulting image quality and imbuing it with certain styles \cite{taxonomy},
TTI models have specific failure modes that we review in the following section.%
%
%
%
\subsection{Failure Modes in Generative Models}
    TTI outputs frequently diverge from user intent due to the limitations of prompting as an interface \cite{Morris}.
One reason is semantic ambiguity. Polysemous terms (such as \textit{``bank''}, \textit{``bat''}, or \textit{``ball''}) can yield images that reflect unintended word senses, creating a mismatch between the user’s mental model and the generative model's output \cite{white2022schrodingersbatdiffusionmodels,10.1016/j.ijhcs.2024.103375}.
Users may intentionally use polysemous and ambigous terms in their creative practice. So-called ``magic terms'' -- lexical additives intentionally designed to induce randomness or probe latent failure boundaries -- 
may result in 
unpredictable or uncanny imagery \cite{taxonomy,maldaner2025miragemultimodelinterfacereviewing,2403.12075.pdf}.



\citeauthor{3491102.3501825.pdf} presented a manual analysis of failure modes in text-to-image models \cite{3491102.3501825.pdf}, identifying recurring breakdowns including underspecified prompts, overloaded terms, and model over-reliance on stereotypical associations. Their findings highlight that many failures stem from subtle mismatches between user phrasing and model priors. This research was later extended to image prompts. \citeauthor{3527927.3532792.pdf} categorized sources of model misinterpretation and proposed design strategies to expose latent failure cases \cite{3527927.3532792.pdf}. These studies contribute to a broader understanding of how prompt formulation interacts with failure patterns in TTI systems.
Failure modes are relevant in interfaces for reviewing and auditing TTI models \cite{maldaner2025miragemultimodelinterfacereviewing,2507.17922.pdf}.
    \citeauthor{2403.12075.pdf}  presented Adversarial Nibbler, a system for crowdsourcing the collection of adversarial examples \cite{2403.12075.pdf},
    and \citeauthor{maldaner2025miragemultimodelinterfacereviewing} presented Mirage, a system for crowdsourced auditing of generative models \cite{Text-35672-1-2-20241013.pdf,maldaner2025miragemultimodelinterfacereviewing}.


Beyond adversarial prompts, training artifacts can manifest as bizarre text--image associations.
    As these pairs have been drawn from the internet, they can include incorrect assignments or mismatched correspondences \cite{feng-etal-2023-uncovering}.
    \citeauthor{daras2022} found ``strange'' associations between special terms in prompts and concepts in generated images \cite{daras2022}. They speculated that these associations in TTI models constituted a ``secret language''.
\citeauthor{williams2024drawlunderstandingeffectsnonmainstream} investigated the effects of non-mainstream dialects in TTI \cite{williams2024drawlunderstandingeffectsnonmainstream}, finding that minimal changes in prompts (including dialect) can manifest biases in TTI models. 
Community reports of special tokens in language models (and by extension, in TTI models), such as \textit{``SolidGoldMagikarp''}, highlight spontaneous `glitch' phenomena that can arise without adversarial intent~\cite{LessWrong,SolidGoldMagikarp}.

In addition to semantic and training-related failures, TTI models exhibit high stochasticity, introducing a form of variability that itself constitutes a failure mode. Identical prompts can produce substantially different images across runs, complicating both user expectation and scientific reproducibility.
Strategies such as sampling multiple random seeds and aggregating outputs have been proposed to mitigate this variability \cite{Liu2022}, and studies show that seed choice alone can 
influence the emergence of rare or novel concepts \cite{samuel2023generatingimagesrareconcepts}.
Failures of prompt recognition -- where the model fails to parse novel or unsupported inputs and defaults to placeholder imagery -- remain underexamined.
Our work targets this gap by presenting a systematic approach for eliciting, identifying, and classifying default-image failures in 
Midjourney.%
%
%

\subsection{Detecting Mode Collapse in Text-to-Image Models}%
%
Mode collapse is a failure mode in generative models where the model produces less diverse outputs than expected.
It was first noted by \citeauthor{NIPS-2014-generative-adversarial-nets-Paper.pdf} in Generative Adversarial Networks (GANs) \cite{NIPS-2014-generative-adversarial-nets-Paper.pdf}.
Default images share characteristics with mode collapse: both involve gravitating toward a limited set of outputs and occur only in response to specific prompts.

Various techniques for detecting and mitigating mode collapse in TTI models have been developed and applied to GANs and diffusion models.
    \citeauthor{978-3-031-90341-0_20} studied early detection of mode collapse through loss monitoring \cite{978-3-031-90341-0_20}.
Other techniques include
    multi-generator approaches
    (e.g., MTC-GAN \cite{Zhang20217789}),
    regularization techniques
    (e.g., ICCR \cite{10315510}),
    contrastive learning
    (e.g., MCL \cite{s40747-022-00924-1.pdf}),
    and novel architectures
    (e.g., style-guided discriminator \cite{Choi_Kim_Song_2022}).
\citeauthor{1910.11626.pdf} used semantic segmentation networks to compare the distribution of objects in generated images with the target distribution to identify omitted classes 
    \cite{1910.11626.pdf}.
    \citeauthor{10334839} developed Connective Novelty Detection, a method that uses an autoencoder and a binary classifier to detect out-of-distribution samples. By creating negative instances close to the positive class distribution and combining real and generated samples, the method improves the robustness of detecting mode collapse \cite{10334839}.
    \citeauthor{Gong2023} used Feature Distribution Divergence Minimization to minimize the divergence between the distributions of real and generated features, reducing the learning pressure on the discriminator and addressing mode collapse without requiring prior knowledge or manual design \cite{Gong2023}.
\citeauthor{Duym_2025_WACV} developed the Mode Collapse Entropy (MCE) score \cite{Duym_2025_WACV} to overcome weaknesses of previous metrics in real-world scenarios.

These approaches are applied at training time and assume full access to either the model or training data, or both, making them impracticable for use in black-box models, such as Midjourney,
where the training data is not known and the model is not directly accessible.
This is particularly important in the context of TTI generation, as the opacity of training image datasets raises questions concerning data provenance, diversity, quality, and representativeness.

Several inference-time methods have been proposed for detecting mode collapse in black-box models.
Statistical, sampling-based tools can be used to visualize and quantify intra-mode collapse without access to training data or model parameters, providing a practical approach for black-box diagnosis and calibration \cite{Wu2021}.
Inversion-based inference perturbation (IIP) alters the inference process to detect memorized images, using unconditional DDIM inversion to extract latent codes and identify images exhibiting collapse-like behavior \cite{Jiang202590485}.
EncoderMI leverages membership inference against contrastive learning–based image encoders, exploiting their overfitting tendencies to reveal whether inputs belong to the training set, which can indirectly indicate mode collapse \cite{Liu20212081}.
Binary noise guidance learning (BnGLGAN) introduces distribution-guided noise reconstruction to stabilize image translation and mitigate collapse \cite{Zhang2024}.

While these techniques require careful statistical modeling or inference perturbation, we adopt a simpler approach. 
We exploit the properties of default images -- visually similar outputs generated from dissimilar prompts -- to identify mode collapse in black-box TTI models (Section \ref{sec:method:comp-analysis}).%
\subsection{Experiments and User Studies on TTI Quality and Satisfaction}%

Evaluations of TTI models often rely on automatic metrics and curated benchmarks, yet these measures are known to misalign with human perception \cite{2203.06026,otani2023verifiablereproduciblehumanevaluation}.
    The Fr{\'e}chet Inception Distance (FID), for instance, can yield scores that disagree with user judgments of image quality and prompt alignment, limiting its utility as a sole evaluation criterion \cite{2203.06026,2403.11821,otani2023verifiablereproduciblehumanevaluation}.
Complementary metrics -- such as CLIPScore \cite{CLIPScore} -- have been proposed to capture semantic fidelity, but also exhibit gaps when compared to subjective assessments \cite{VQAScore}.
Recognizing these shortcomings, \citeauthor{russo-2022-creative} suggested criteria for a creative TTI benchmark  incorporating insights from linguistics, cognitive psychology, and art theory to better capture human notions of creativity and relevance \cite{russo-2022-creative}.
More recently, the Holistic Evaluation of Text-to-Image Models (HEIM) benchmark systematically assessed 26 state-of-the-art TTI generators across twelve dimensions (ranging from aesthetics and originality to bias and robustness) using a combination of automated measures and human ratings, providing a richer, multidimensional view of model performance \cite{2311.04287}.

Human‐centered studies 
bridge the gap between automatic scores and actual user experience.
    The Pick-a-Pic dataset includes over half a million human preference annotations, enabling the training of PickScore, a model predicting human rankings of generated images with superhuman accuracy, and suggesting a more reliable protocol for evaluating TTI outputs \cite{NEURIPS202373aacd8b}.
Researchers extended this work with ImageReward, a preference‐based reward model from 137,000 expert comparisons. Fine-tuning diffusion models against this reward function produces images that users consistently rate as higher quality \cite{xu2023imagerewardlearningevaluatinghuman}.
\citeauthor{otani2023verifiablereproduciblehumanevaluation} highlight the lack of verifiable, reproducible human evaluation in TTI research and propose a standardized, crowdsourced protocol \cite{otani2023verifiablereproduciblehumanevaluation}. Their results reveal that automatic metrics frequently diverge from human judgments, underscoring the need for rigorous user studies.
In interactive settings, tools, such as PromptAid~\cite{mishra2025promptaidpromptexplorationperturbation} and  PromptCharm~\cite{3613904.3642803}, helpedparticipants craft prompts yielding images better aligned with their intent, with users reporting increased satisfaction and perceived control over output quality. Multi-turn guidance systems, such as DialPrompt~\cite{2408.12910}, achieve significant gains in both image quality ratings and user‐centricity scores compared to baseline prompt‐engineering interfaces.
Finally, exploratory work in teacher education demonstrates how practitioners reflect on prompt strategies and generator behaviors in art and design, revealing both learning affordances and cognitive challenges when integrating TTI tools into creative pedagogy~\cite{Ringvold}.

Together, this prior work illustrates the complex interplay between objective metrics, human preference, and interaction design in assessing and improving TTI systems.%

\section{Method}%
\label{sec:method}
We present the first investigation into default images,
acknowledging that the complexity 
of TTI models limits us to a tiny glimpse of the latent space and its default images.
Our approach involves systematic prompt creation (Section \ref{sec:method:prompts}), 
manual image generation experiments and ablation studies to probe the latent space (Section \ref{sec:method:experiments}),
identification of default images using affinity diagramming (Section \ref{sec:method:affinitydiagramming}), a user study on user satisfaction with default images (Section \ref{sec:method:surveys}), and a large-scale computational analysis of Midjourney images (Section~\ref{sec:method:comp-analysis}).%
\subsection{
Input Prompts Likely Triggering Default Images}%
\label{sec:method:prompts}%
In our initial exploration of the phenomenon of default images,
we explored three categories of prompts that 
could result in a conceptual mismatch between user expectations and the resulting image.
%
We first tried to `confuse' the model with polysemous words or prompts outside the training data (e.g., e-mail addresses), but the rate of default images was too low.
We then tried prompts that were likely not recognized by the model and lacked practical purpose, including (but not limited to):
    contextually misleading phrases (e.g., \textit{Einstein’s relativity theology}), 
    random character strings, emojis, and special characters (e.g., \textit{aölkjnewjbtn} and \textit{\%}),
    lyrics and misheard lyrics,
    and abstract and metaphorical concepts that may not invoke strong imagery (e.g., \textit{absolute ownership} and \textit{extreme aptitude}).
We excluded prompts in this category from further analysis because, although they often trigger default images, they are not intended to produce coherent visual content.
These prompts often function as stress tests or purposeful creative experiments rather than artistic image generation attempts.
While these prompts remain important for future investigations into model boundaries and creative prompting strategies, in this work, we focus on default images arising from realistic prompting scenarios, where users aim to generate meaningful images (instead of default images).

\label{sec:prompts}
We explored prompts that involve both unrecognized terms and practical use cases, a combination
likely found in real world use, but that still challenge the TTI model.
These include:%
\begin{itemize}%
    \item[A1:]
Rare or 
uncommon names, such as fantasy names, 
(e.g., \textit{Kirk Kirkson} and \textit{Shelan Creswell})
    \item[A2:]
Corrupted words or altered words (e.g., \textit{tractoajrte})
    \item[A3:] 
Web addresses (e.g., \textit{\nolinkurl{www.facebook.com}} and \textit{\nolinkurl{www.quora.com}})
    \item[A4:] 
Words from low-resource languages, like A4.1 Finnish (e.g., \textit{hevonen} [horse]), and A4.2 Tagalog, an Austronesian language spoken mainly in the Philippines, (e.g., \textit{gubat} [forest])
    \item[A5:]
Glitch tokens \cite{SolidGoldMagikarp} (e.g. \textit{rawdownload})
    \item[A6:]
Syllables or abbreviations (e.g., \textit{BCD} and \textit{YMCA})%
\end{itemize}%

We manually created a set of 130~prompts based on the six categories, A1--A6.
We created prompts in category A2 by introducing altering known words with slight variations.
We selected the two low-resource languages (A4.1 Finnish and A4.2 Tagalog) for their language-native words that are 
    not borrowed from other languages, so they 
do not cognate with widely known foreign languages.
We collected the glitch tokens from related literature \cite{SolidGoldMagikarp,SolidGoldMagikarp2,SolidGoldMagikarp3}.
The prompts are in Appendix~\ref{appendix:tokens}).
%

To assess the likelihood that the selected prompts had been encountered during model training, we checked each prompt on \href{https://www.haveibeentrained.com}{haveibeentrained.com}, a search interface for over 5.8 billion image-caption pairs.
The tool helps users determine whether a specific term (e.g., their name) appears in large-scale datasets, such as LAION \cite{Laion}. These datasets are commonly used in the TTI model training \cite{Laion,NEURIPS2022a1859deb}. We sought prompts that were absent from -- or underrepresented in -- the training data, increasing the likelihood that the resulting generated imagery would involve default or fallback behavior.
Our inclusion criteria were twofold: 1) the prompt should plausibly occur in real-world use, and 2) it should pose a challenge to the model due to low training data coverage.
%
We describe our image generation process next.%

\begin{figure*}[!t]%
    \centering
    \includegraphics[width=.9\textwidth]{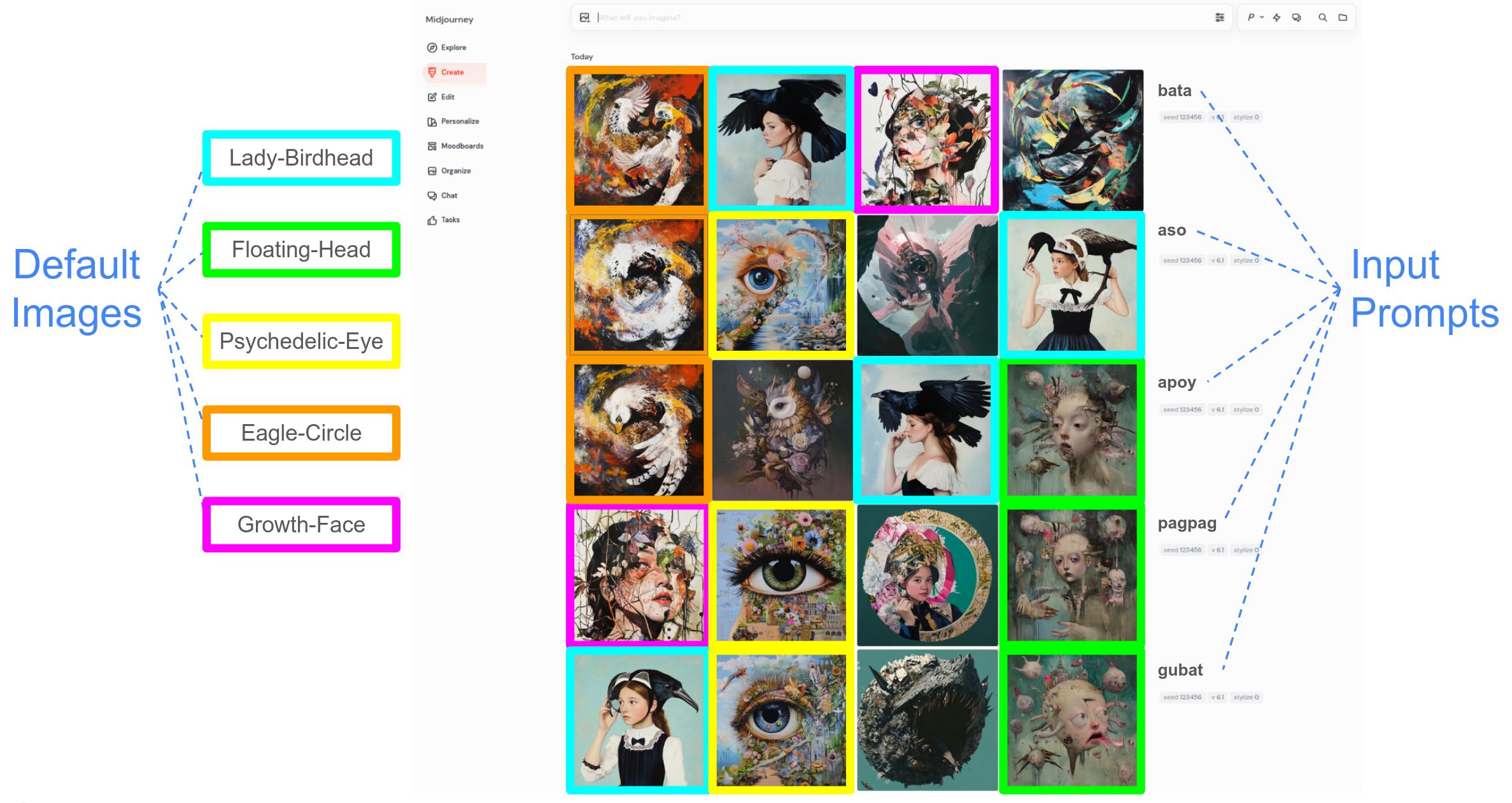}
    \caption{%
    \rev{%
    Default–prompt relationships in Midjourney.
    A screenshot illustrating how diverse input prompts (e.g., the Tagalog words \textit{bata}, \textit{aso}, \textit{apoy}) yield overlapping clusters of recurring default images.
    }%
    }
    \Description{A screenshot of the Midjourney page, with default images in different prompts highlighted. The figure demonstrates how different default images emerge from different prompts.}
    \label{fig:screenshot}
\end{figure*}%

\subsection{Image Generation Experiments and Ablation Studies}%
\label{sec:method:experiments}%
We used Midjourney's 
    webpage\footnote{\href{https://www.midjourney.com/imagine}{www.midjourney.com/imagine}} (see \autoref{fig:screenshot}) to execute the experiments, using fixed parameters listed in \autoref{tab:parameters}.
For each of the 130~prompts in the six input categories (A1--A6), we ran 1~trial, resulting in four images per prompt, for a total of \ok{520}~images.
We then visually analyzed the images, as described in Section~\ref{sec:method:visualgroundedtheory}.

\begin{table}[htb]%
\centering
\small
\caption{Image generation parameters used in our manual experiments}%
\label{tab:parameters}%
\begin{tabular}{lll}
\toprule
    Parameter & Value & Description \\
\midrule
    model  &  6.1 & setting the model version \\
    aspect ratio & 1 (square) & 
        standardizing the format of all generated images \\
    stylize &  0 & controlling the artistic flair\\
    chaos & 0 & controlling having surprising random elements\\
    weird & 0 & controlling having quirky and unconventional elements\\
    seed value & 123456 &  ensuring consistency and reproducibility\\
\bottomrule
\end{tabular}
\end{table}%

As a sensitivity analysis to test the robustness of default images, we conducted five small-scale ablation studies.
We investigated the effect of different seed values on the default image (Section \ref{sec:ablation:seeds}).
We explored whether default images could be altered with prompt modifiers \cite{taxonomy} (Section~\ref{sec:ablation:modifiers}).
We then examined default image behavior within larger prompts, including known terms (Section \ref{sec:ablation:largerprompts}) and combinations of prompts known to trigger default images (Section \ref{sec:ablation:combos}.
In our last study, we investigated 
default images in different model versions
(Section \ref{sec:ablation:models}).%
%
%
%
%
\label{sec:method:visualgroundedtheory}%
%
%
%
%
%
%
%
%
%
%
%
%
\subsection{Default Image Identification}%
\label{sec:method:affinitydiagramming}%
To identify default images in the set of generated images, we used affinity diagramming (or KJ Method) \cite{KJMethod}. It is a qualitative method that organizes data into groups based on perceived similarity. The method is commonly employed to
    categorize complex data.
The first two authors collaboratively grouped images on a digital canvas based on their perceived similarity.
During the session, the two authors discussed their rationale and iteratively refined the groups.
Simultaneously, the two authors developed descriptive labels for the images (e.g., ``Lady-Birdhead''), enabling them to refer to these images by name.
They continued this collaborative process
until all images were either categorized as default images (those with strong similarities across multiple images created with different prompts) or not.
Identifying similarities between images was often straightforward despite subtle variations, as visible 
in \autoref{fig:examples-variations}. 
%
%
\subsection{Online User Study}%
\label{sec:method:surveys}%
We used crowdsourcing through Prolific,\footnote{www.prolific.com} an online human subject pool, to 
explore participant perception of default images, 
%
%
%
specifically, the impact of default images on user satisfaction.
We used an online
a questionnaire presented participants with four images, including the original prompt that created the image.
Unknown to the participant, for two of the images (Q1 and Q2), we replaced one term in the prompt with a term identified as causing default images.
    For instance, in Q1, we replaced \textit{`raven'} in the prompt \textit{``photoshoot of a woman wearing a white off-shoulder dress with a black raven on top of her head''} with \textit{'Greagoft'}.
    For Q2, we replaced ``leaves and flowers'' with ``Shuttastfr''.
    With this replacement, we simulated the situation where a word is unknown to the generative model.
    We showed participants the original prompt and the image resulting from the modified prompt, simulating the mismatch in expectations that an encounter with a default image can produce.
For Q3, we used an image that matched the descriptive prompt, without replacements.
For Q4, we used a prompt that was unknown to the model and, thus, the accompanying image was a default image.
The four prompts and images are in \autoref{fig:survey2-results}.
We asked participants to rate their satisfaction with each image, given the prompt, on a scale from 1 (Strongly Dissatisfied) to 7 (Strongly Satisfied).
Participants received {\pounds}6.29 per hour, and the median time to complete the survey was 3~min and 49~seconds.




\subsection{Computational Analysis of Midjourney Default Images}%
\label{sec:method:comp-analysis}%
We extended our study to real-world data on Midjourney to examine whether the phenomenon of default images exists at scale and in real-world usage.
We applied a similar approach as in our initial manual study, and used the definition of default images -- visually similar images created from dissimilar prompts -- to computationally identify clusters of default images in Midjourney data.%
%
%
\subsubsection{Data collection}%
\label{sec:method:data-collection}%
%
We collected image-prompt pairs from Midjourney with DiscordChatExporter,\footnote{https://github.com/Tyrrrz/DiscordChatExporter} a tool to retrieve Discord messages. We downloaded a sample from the Midjourney Discord channels `general-2', `general-3', `general-18', and `general-19', starting from the earliest messages.
We included only messages originating from the Midjourney Bot with image attachments.
We further excluded images with non-square aspect ratio,
image prompts (i.e., prompts containing \textit{`https://'}),
panned images,
zoomed images,
variations,
image inpainting,
and upscaled images.
The complete filter settings are as follows:
\begin{quote}
    \small
    \textit{
    {-}{-}media {-}{-}reuse-media {-}{-}filter "from:'Midjourney Bot' \& has:image \& -'https://' \& -'—ar ' \& -'\textbackslash-\textbackslash-ar' \& -'Upscaled' \& -'Zoom Out' \& -'Pan Right' \& -'Pan Left' \& -'Pan Up' \& -'Pan Down' \& -' \textbackslash- Image \#1 \@' \& -' \textbackslash- Image \#2 \@' \& -' \textbackslash- Image \#3 \@' \& -' \textbackslash- Image \#4 \@' \& -'\textbackslash-\textbackslash-draft' \& -'\textbackslash-\textbackslash-sref' \& -'Variations by' \& -'Variations (Region) by' \& -'Variations (Strong) by'
    \& -'Remix by'
    \& -'Remix (Strong) by'
    \& -'Remix (Subtle) by'
    \& -'\textbackslash-\textbackslash-p'" {-}{-}before 2025-04-03%
    }%
\end{quote}
All downloaded images are panels of four square images, which we split into four separate images with a script.
We excluded drafts, style references, and personalization.
We also excluded model version 7 (released 2025-04-03), because of its personalization feature.
Because we applied these filtering criteria directly during the scrape, we do not know the total number of images processed and excluded.

From the resulting \ok{6,302,304} prompts collected,
we removed
exact duplicates. We also excluded
prompts larger than 25 characters (counting multi-byte characters as one character) as all of the prompts in our manual study are less than 25 characters in length.
We cleaned the prompts by removing any remaining parameters (such as \textit{{-}{-}relax}, \textit{{-}{-}weird}, and \textit{{-}{-}stylize}) and applied FastText’s language identification model (lid.176.bin) to detect the language in the prompts.
%
%
We took the Midjourney version from explicit parameters (such as \textit{--v 5.1}) or, if the parameter was not present, inferred it from model release dates. The inferred version is only an estimate, because Midjourney sometimes tests new models before making them the default model.
We did not perform filtering for copyrighted content or sensitive terms in prompts and instead relied on Midjourney's built‑in moderation to limit such cases.%

\subsubsection{Dataset description}%
\label{sec:dataset}

\begin{figure}[!htb]%
\centering%
    \includegraphics[width=.4\linewidth]{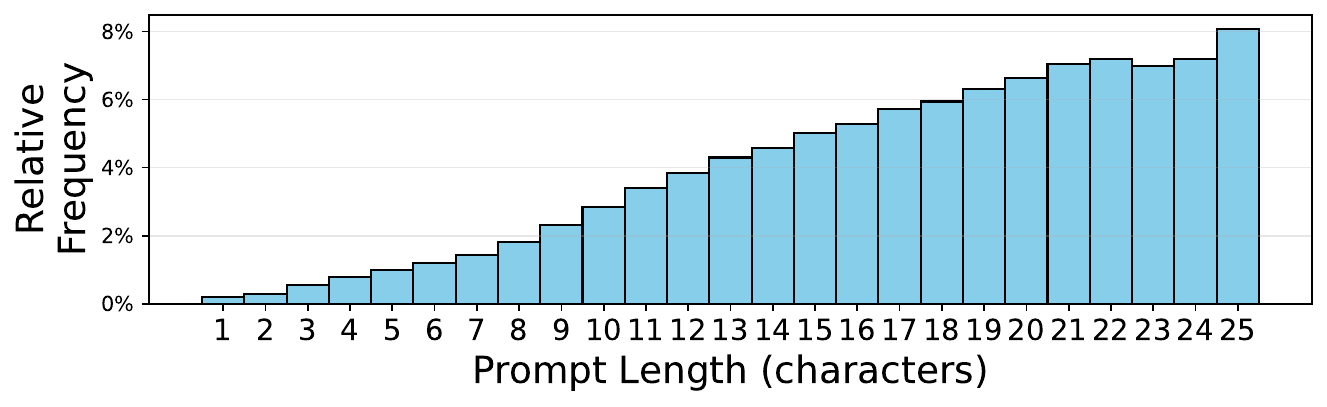}%
\caption{
\rev{Prompt-length distribution. Shows the relative frequencies of prompt lengths (in characters) in the dataset.
}%
}%
\Description{The distribution of prompt lengths shows that very short prompts are rare in the dataset. Prompts with 18 or more characters make up about 50\% of the dataset.}%
\label{fig:prompt-length}%
\end{figure}

The dataset contains
    \rev{189,432} prompts,
from July 9, 2022 to March 17, 2025.
Each prompt produced exactly four images, for a total of 
    \rev{757,728} images.
This includes 
     \rev{29,872}  images for Midjourney version 1.0,
     \rev{92,496}  images for v2.0, 
     \rev{173,468} images for v3.0,
     \rev{109,672} images for v4.0,
     \rev{64,092}  images for v5.0,
     \rev{67,248}  images for v5.1,
     \rev{60,300}  images for v5.2,
     \rev{108,868} images for v6.0,
     and
     \rev{51,712}  images for v6.1.%
The mean prompt length is \rev{17.5} characters \rev{($Median=18$, $SD=5.5$, $Q1=14$, $Q3=22$)}.
%
%
Some prompts include special tokens and punctuation,
    although single character prompts are rare (see \autoref{fig:prompt-length}).
%
%
The prompt languages include 
English  (\rev{47.2\%}),
Italian  (\rev{5.4\%}),
Romanian (\rev{4.0\%}),
French   (\rev{3.1\%}),
and many others.
The columns included in our dataset are described in \autoref{tab:datasetcolumns} in Appendix~\ref{appendix:dataset}.

%
%
%
%
\subsubsection{Data analysis}%
\label{sec:method:data-analysis}%
%
To analyze the dataset, 
we follow the same approach as our manual study.
We first cluster visually similar images and then keep only clusters with dissimilar prompts.

We experimented with different solutions to identify visually similar images, including training a joint image-prompt embedding with a contrastive learning loss, as well as image similarity measures, such as 
cosine similarity between pixel vectors,
LPIPS 
    \cite{1801.03924.pdf},
SSIM~\cite{1284395},
and MS-SSIM~\cite{1284396}.
We found the CLIP model (ViT-L/14) \cite{CLIP,VIT} produced the most visually coherent clusters for our use case.
Using CLIP, we transformed the images into embeddings (i.e., numeric vectors), which we precomputed locally and then transferred to a cloud server.
We then computed the full pairwise cosine similarity matrix between embeddings to identify visually similar images.
Because the number of unique pairs grows quadratically with the number of images, the computation and storage scale as $O(N^2)$. In practice, we exploit symmetry by evaluating only the upper triangle and mirroring. To fit in memory, we computed matrices separately for each major Midjourney version.
All computations were performed on a cloud server with 96~vCPUs and 1~TB RAM.

To find \rev{clusters} among the visually similar images, we employed a clustering algorithm.
Again, we experimented with different solutions, such as connected component clustering and HDBSCAN \cite{HDBSCAN}, but settled on simple agglomerative hierarchical clustering \cite{10.1002/widm.53} with a threshold of 0.1 on cosine distances between normalized CLIP embeddings. This threshold controls cluster formation by stopping merges once the cosine distance between clusters exceeds 0.1, thereby grouping images with high visual similarity.
We
set the minimum cluster size to $n=3$ images.
From visual inspection, we noted that clusters were individually tight but sometimes similar to each other \rev{across clusters}, a common artifact of hierarchical clustering with a fixed distance threshold.
To address this, we implemented a post-processing merge of clusters based on their centroid similarity with a threshold of 0.9.

This resulted in many clusters of visually similar images.
However, they contained images that were often created from 
    lexically (e.g., \textit{dragons} and \textit{a grim dragon}) or semantically (e.g., \textit{dog} and \textit{labrador}) similar prompts.
Default images, however, arise from lexically and semantically dissimilar prompts.
%
%
To address this, we developed a two-stage rule-based filtering process, first filtering out clusters with high lexical overlap in a clusters' prompt words, and then filtering clusters with high semantic similarity.
\rev{%
The first filter counted the number of appearances of each word in prompts within the cluster. If any word occurred in more than 50\% of the prompts within the cluster, we excluded the cluster from further analysis.
Formally, let a cluster contain \(M\) prompts, and let \(V\) be the set of all distinct words appearing in those prompts. 
For each word \(w \in V\), let \(\mathrm{freq}(w)\) denote the number of prompts in the cluster that contain \(w\). 
We then compute the lexical–overlap score
\[
L = \max_{w \in V} \frac{\mathrm{freq}(w)}{M}
\]
and exclude the cluster if $L \ge 0.5$.
This simple lexical-overlap filter removed prompts where images were similar due to the same word occuring in prompts. It did, however, not filter visually similar images resulting from semantic similarity in prompts (such as the prompts \textit{labrador} and \textit{dog}).
This was addressed with a second filter in which
}%
%
we assessed the intra-cluster semantic similarity of prompts.
We used sentence transformers (all-MiniLM-L6-v2) and calculated a numeric score representing the intra-cluster semantic similarity -- the average pairwise similarity between all prompt embeddings within a cluster.
After computing over 100~clustering solutions with different parameters and visually inspecting the results, we found that clusters with semantic similarity scores over {0.3} often had prompts semantically related to the same topic 
    not covered by our lexical filter.
    For this reason, we excluded any clusters with semantic \rev{similarity} score \rev{$>$ 0.3}.
\transition{This two-staged filtering by \rev{lexical diversity and semantic similarity} helped us identify \rev{candidates of} default image clusters in the dataset.}


\section{Results}%
\label{sec:results}%
%
\subsection{Default Images}%
\begin{figure}[!htb]%
    \centering%
    \includegraphics[width=\linewidth]{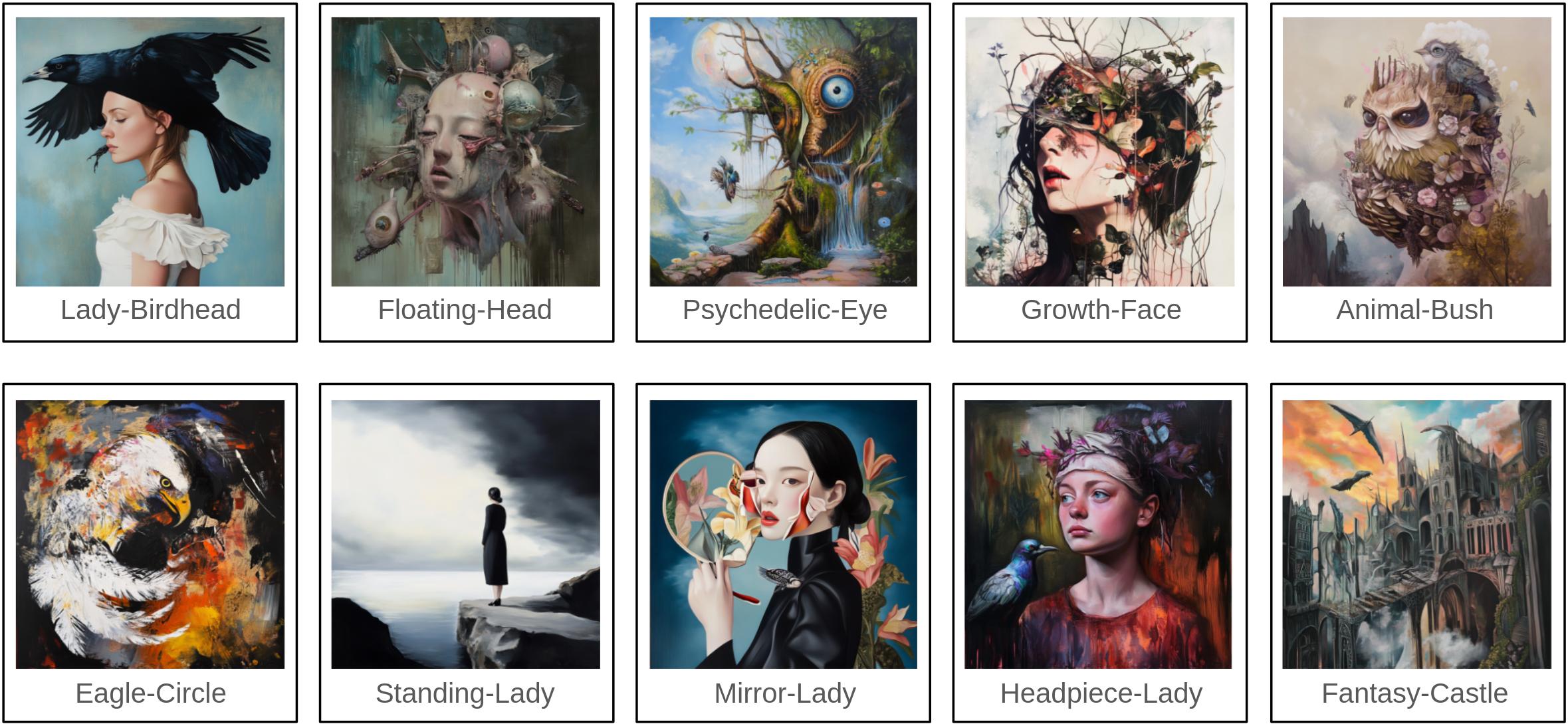}%
    \caption{
    \rev{%
Set of default images. Shows the set of default images identified in our manual study, along with the descriptive labels we assign for future reference.
    }%
    }%
    \Description{The figure depicts 10 default images identified with affinity diagramming. Many of these images appear in other figures in this paper, with slight variations.}%
    \label{fig:Defaultimagesfromtestruns}%
\end{figure}

\begin{figure*}[htb]%
    \centering%
        \includegraphics[width=.75\linewidth]{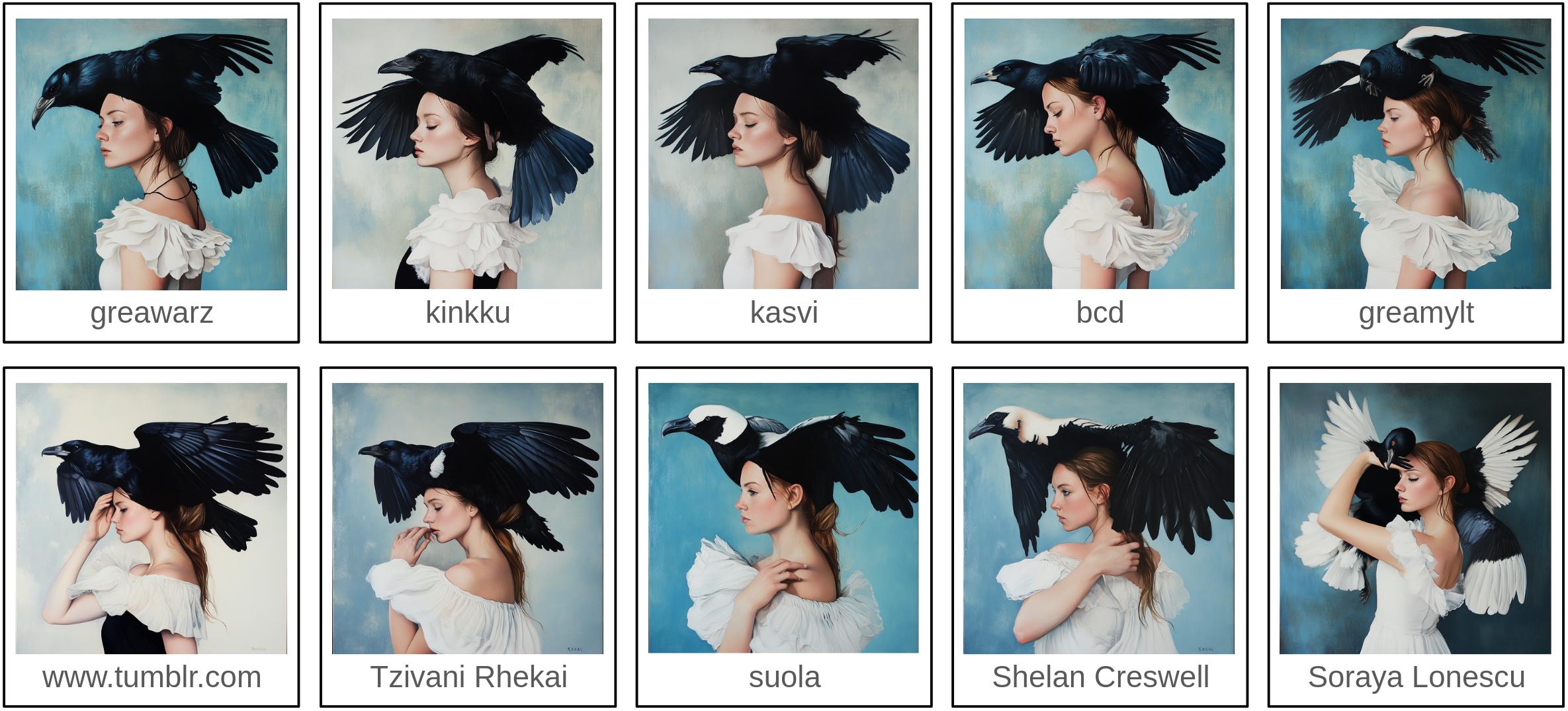}%
    \caption{
    \rev{%
    Same default image from dissimilar prompts. Demonstrates that diverse and semantically unrelated prompts still yield the same default image in Midjourney v6 and v6.1, with only minor variations.
    }%
    }%
    \Description{The figure depicts some of the variations encountered of the default image `Lady Birdhead'. This image shows a woman with a black bird on her head. In all images, the woman is wearing a white dress and faces left. The are variations in her clothing, in her haze, in the way she holds her arms, and in the bird.}%
    \label{fig:examples-variations}%
\end{figure*}

\begin{figure*}[htb]
    \centering
    \includegraphics[width=.95\textwidth]{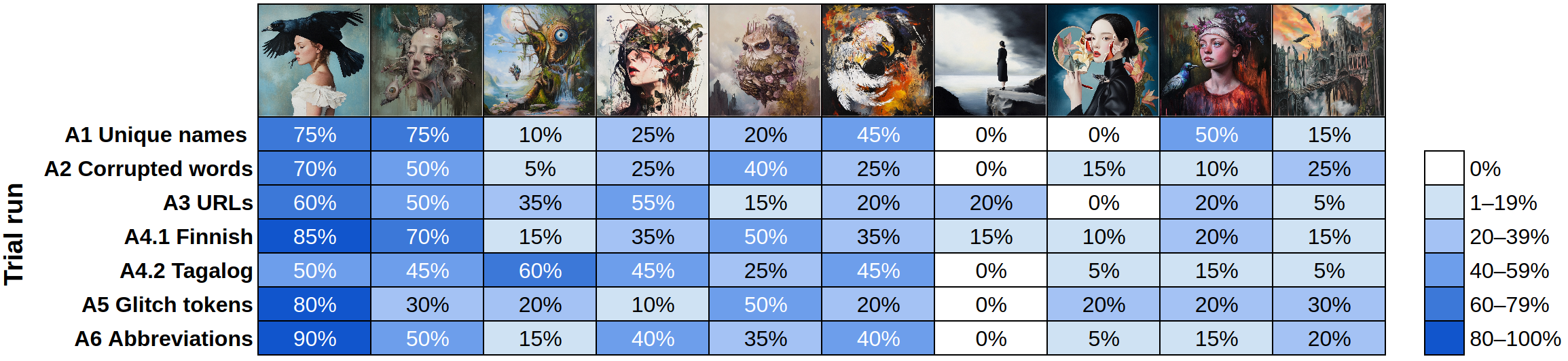}
    \caption{
    \rev{%
    Default-image frequencies across test sets.
    Shows how often each default image appears in our six prompt groups (A1–A6), revealing consistent patterns across disparate prompt types.
    }%
    }%
    \Description{The figure shows the frequency with which different default images, as identified with affinity diagramming, have appeared in our manual experiments. The default images named 'Lady Birdhead' appeared most often.}
    \label{fig:frequencies}
\end{figure*}

\begin{figure*}[!thb]%
    \centering%
        \includegraphics[width=\textwidth]{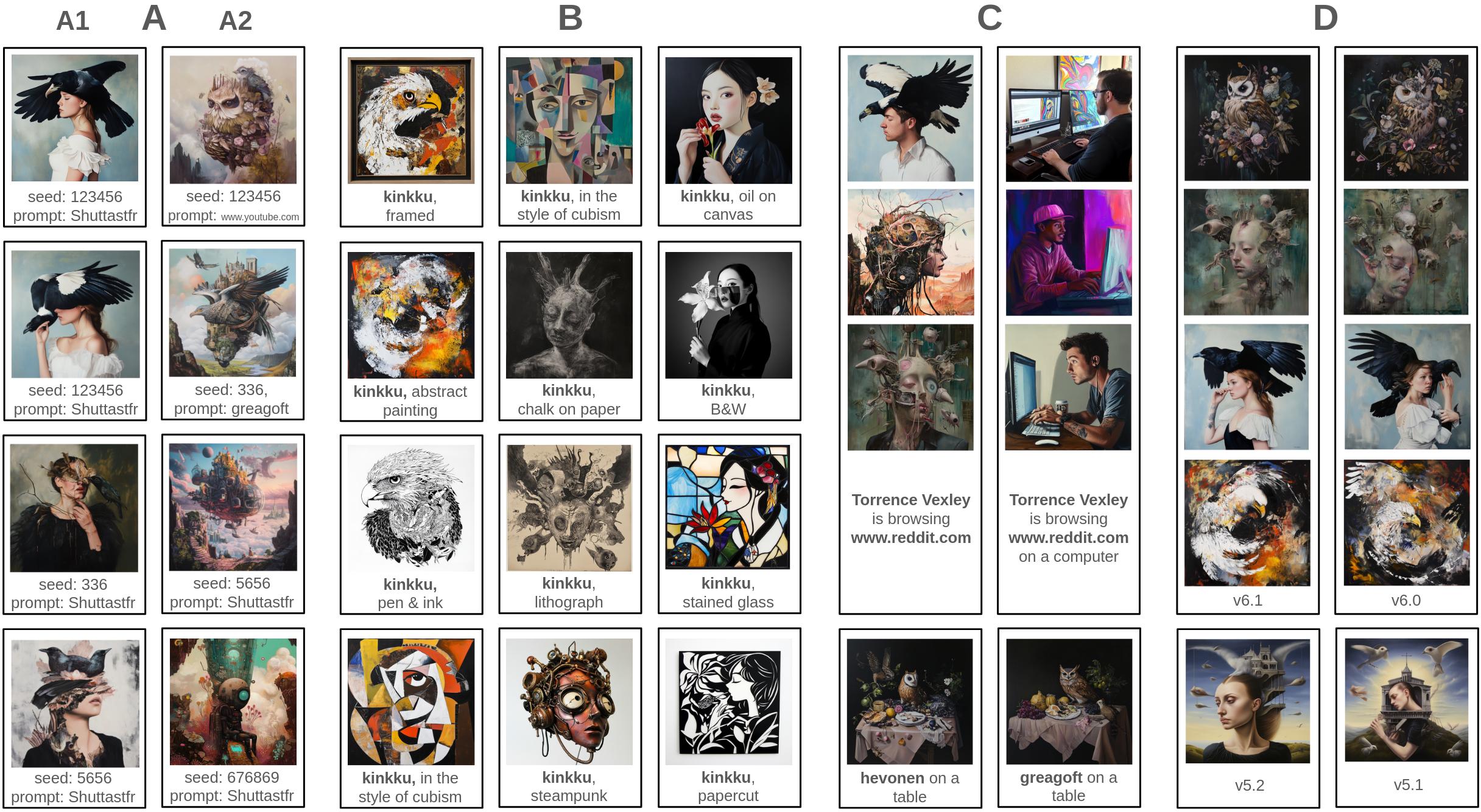}%
    \caption{%
    \rev{%
Ablation results for default-image behavior.
Shows the effect of seed choice, style modifiers, known-term dominance, and model version on default images.\\
A: Default images remain consistent across varied seeds and prompts.\\
B: Style modifiers change appearance but preserve the underlying default-image 
motif.\\
C: Known concepts override unknown tokens, yielding non-default images.\\
D: Some Midjourney versions generate visually similar defaults from the same prompt.%
    }%
    }%
    \Description{The figure depicts the results of our ablation studies, demonstrating how default images show variations with different seed values, style modifiers, model versions, or when used as part of larger prompts.}%
    \label{fig:Varyingseed}%
    \label{fig:model-comparison}%
    \label{fig:Prompts1}%
\end{figure*}

%
\autoref{fig:Defaultimagesfromtestruns} showcases the ten default images 
identified \rev{with affinity diagramming} in our manual study, based on the set of input prompts from Section~\ref{sec:method:prompts}.
Some default images occurred more frequently, as depicted in \autoref{fig:frequencies}.
In this figure, A1--A6 denote the different test sets, comprising 130 prompts (Section \ref{sec:prompts} and Appendix \ref{appendix:tokens}).
\autoref{fig:examples-variations} showcases how the same default image (in this case, `Lady-Birdhead') appears in response to different prompts, with only minor visual variations.
Due to Midjourney's internal diversity sampling mechanism, the relationship between prompt and default image is many-to-many: one prompt can produce multiple different default images, and the same default image can appear in response to multiple different prompts (cf. \autoref{fig:screenshot} and \autoref{fig:model-comparison}B).

The default images we identified exhibit recurring motifs, reflecting the model's aesthetic biases when faced with ambiguous or unrecognized prompts. Human figures and natural elements -- such as birds, clouds, trees, and flowers -- appear as primary motifs, separately or merged into strange creatures.
Human-vegetal and/or human-animal fusions emerge as key subjects in default images, appearing across different atmospheres, from oneiric scenes to mythological metamorphoses.
Some identified default images confirm that similarly to human dreams, ``AI has an uncanny ability to combine seemingly disparate elements into sometimes sinisterly coherent representations'' \cite{PRADA}.  

Although the literature on AI imagery attests that visual expression tends to vary according to specific TTI generators, and some note `default styles' associated with different models \cite{MANOVICH}, the curated series of 10~images presented in \autoref{fig:Defaultimagesfromtestruns} reveals significant aesthetic disparities. While the five images in the lower row are reminiscent of abstract expressionist and surrealist painting, also evoking magical realism and epic fantasy films, the first five images draw from other artistic movements. As shown in figures \ref{fig:Defaultimagesfromtestruns} and \ref{fig:examples-variations}, the female portrait is also an important theme, particularly prone to formal and stylistic derivations. From the solitary woman standing before a romantic landscape -- in which Caspar David Friedrich's painting resonates -- to close-up representations that seem to combine the tradition of portrait painting with lyrical abstraction and contemporary fashion photography, these default images illustrate the hybrid and composite nature of AI-generated pictures. 
\autoref{fig:examples-variations} further emphasizes how AI-generated pictures can condense multiple artistic references into a single visual composition that, similarly to a musical piece, can then unfold in subtle variations. This series reflects the wide dissemination of individual portraits, both through the digitization and online circulation of art collections -- including popular museum masterpieces -- and through the massive production of born-digital pictures that fuel social media.

The intriguing association of a stereotyped female figure and a bird in place of a hat \rev{in \autoref{fig:examples-variations}} poses a twofold question concerning the generation of default images. Apparently, the TTI model was unable to understand the prompt or correctly recognize some of the images it was trained with. Feather hats are a recurring theme in the history of female portraiture and fashion since the Renaissance, from Titian's Portrait of a Young Woman (c. 1536) to Rubens' Portrait of Helena Fourment (c. 1630), from Klimt's The Black Feathered Hat (1910) to Karl Lagerfeld's exuberant designs for Chanel's Spring Collection in 1992. The examples are countless, hence it seems plausible that many of these images were used to train the model. However, there seems to be a fundamental difference between human perception of visual representations of feather hats and the way computer vision `sees' such images. Rather than a dataset bias, what is at stake here is a perceptual bias \cite{OFFERT}. Following a metonymical logic (i.e., taking a part to stand in for the whole), the model takes the part for the whole, assuming that the feathers correspond to the presence of a bird whose species differs across the variations of the default image. This is an insightful demonstration of ``the nomadic existence to which images seem to be condemned within the historical formations of knowledge'' \cite{ALLOA}. 

\subsection{Ablation Studies}%
To evaluate the consistency of the default images generated by Midjourney in response to our set of input prompts, we examined how altering the seed value, prompt, and model version affected the default images.%
\subsubsection{Effects of seed values}%
\label{sec:ablation:seeds}%
We visually analyzed how 
different seed values affect the default images. 
Using different seeds with the same prompt resulted in different default images, partially sharing some of the motifs (\autoref{fig:Varyingseed}--A1).
While there are similarities, one might not associate these images with each other. Thus, the best way to generate default images is by having a fixed seed value for all image generation runs.
\rev{%
The images in \autoref{fig:Varyingseed}--A2 show default image similarities with variations in both seed values and prompts.
}%
\subsubsection{Effects of style modifiers on default images}%
\label{sec:ablation:modifiers}%
We found default images respond to style modifiers just like regular \rev{generated} images.
The same default image motifs reliably arise \rev{when applying different styles.}
    \autoref{fig:Varyingseed}--B depicts how the prompt \textit{``kinkku''}, which triggers default images, can be modified with different styles, like \textit{``in the style of cubism''} or \textit{``oil on canvas''}.%
%
%
%
\subsubsection{Effects of default prompts within larger prompts}%
\label{sec:ablation:largerprompts}%
We experimented with prompts that are known to trigger default images as part of larger prompts to mirror real-world usage. 
As shown in \autoref{fig:Prompts1}--C, a short prompt with unknown terms may generate default images.
The short prompt also produces more varied images because it places fewer constraints on the model.
However, when those same terms are included in longer prompts, the model can generate imagery from any words that it recognizes.
Using longer prompts resulted in images distinct from the default images.%
%
%
%
\subsubsection{Effects of combining multiple default image prompts}\label{sec:ablation:combos} We found that when we combined two of the default image prompts (from either the same or different sets), they would still reliably yield default images. However, \rev{as the number of default image triggering terms increased,} the number of resulting default images appeared to decrease. When we combined all 20~prompts from a set, the default images identified in the main study (see Figure \ref{fig:Defaultimagesfromtestruns}) were rare.%

\begin{figure*}[!thb]%
\centering
    \includegraphics[width=.96\textwidth]{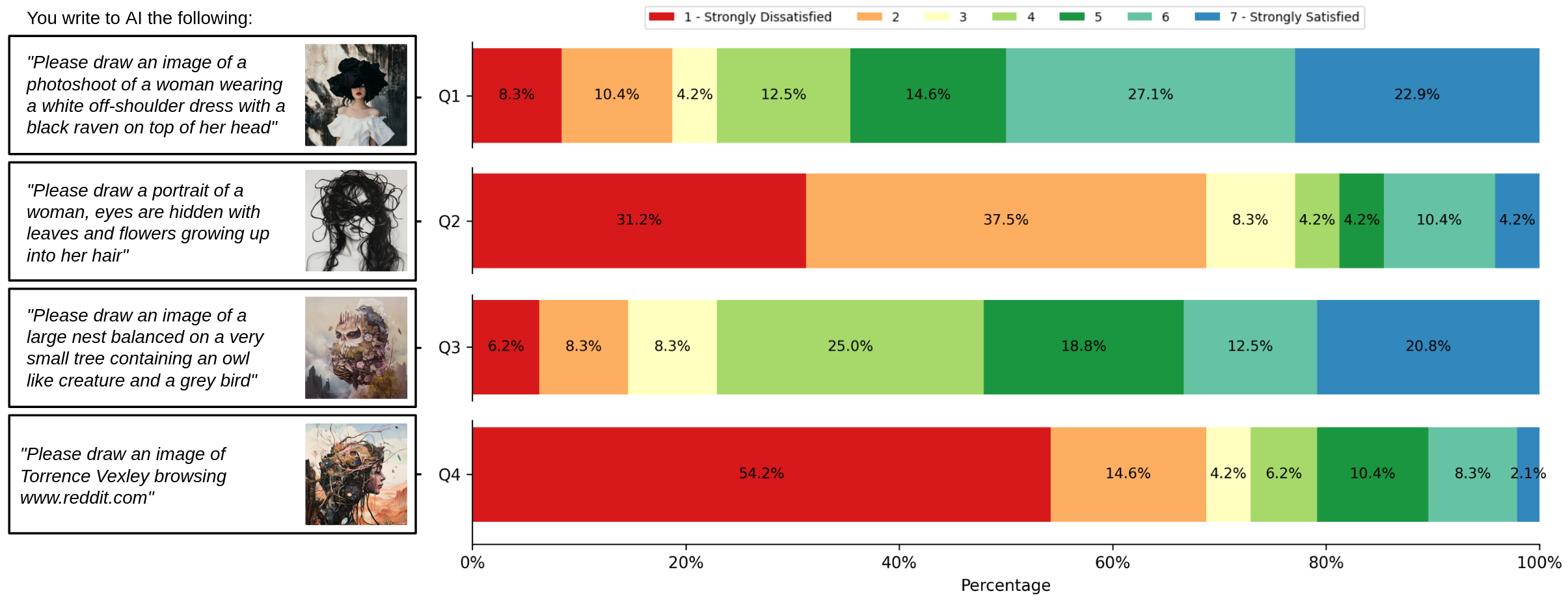}
\caption{
\rev{%
User satisfaction with default-image outputs.
Responses from our online study \ok{(N~=~48)} show that default images can lead to user dissatisfaction.
}%
}%
\Description{Responses from the user study (N=48) show how default images can lead to user dissatisfaction.}
\label{fig:survey2-results}
\end{figure*}%

\subsubsection{Effects of model versions}%
\label{sec:ablation:models}%
Last, we analyzed how different model versions affect the default images.
In our experiments with Midjourney, both model versions 6.0 and 6.1 consistently generated very similar default images with the same prompt, as depicted in \autoref{fig:model-comparison}--D.
Older model versions (e.g., 5.1 and 5.2) generated default images with a more surreal \rev{or abstract} aesthetic, but with similar motifs, including floating objects, human portraits intertwined with natural elements (plants or animals, or both).

\subsection{Default Images and User Satisfaction}%
\label{sec:surveyresults}%
%
\subsubsection{Participants}%
The study participants \ok{(N~=~48)} were between \ok{21 and 68} years old \ok{($Mean = 32.9$, $SD = 9.7$)}
with \ok{27} men and \ok{21} women.
About \ok{52\%} of the participants had completed a Bachelor degree and \ok{23\%} a Master's degree.
Participants were familiar with TTI \ok{($Mean=6.0$ and $SD=1.3$} on a scale of 1--Strongly Disagree to 7--Strongly Agree).%
\subsubsection{Results}%
\autoref{fig:survey2-results} depicts the results of our user study, showing that \rev{default images} 
negatively affect user satisfaction, but not in all cases.%

In Q1, implicitly replacing \textit{`raven'} with \textit{`greagoft'} resulted in only subtle image changes. The image appeared almost as if a raven could be there.
The participants' satisfaction, therefore, was relatively high \ok{($Mean=4.9$, $SD=2.0$)}.
In Q2, where \textit{`leaves and flowers'} was replaced with \textit{`greagoft'}, the divergence between prompt and resulting image was more noticeable and satisfaction was lower \ok{($Mean=2.6$, $SD=1.8$)}.
A Fisher's 
test with continuity correction indicated that 
participants' perception of the change in the resulting image (subtle vs. noticeable) was significantly related to their satisfaction level
\ok{($OR=10.7$, $p<0.05$)}.
\ok{The odds ratio of \ok{10.7} indicated that dissatisfaction (1--3 on the Likert scale) was much less likely when the change was subtle compared to noticeable. In other words, participants were significantly more likely to be satisfied (5--7) when the change in the image was subtle (Q1) rather than noticeable (Q2).}

For Q3, we did not switch or omit any words in the prompt: the image was the exact result of the prompt.
Participants were satisfied with the resulting image \ok{($Mean=4.6$, $SD=1.8$)}.
For Q4, we used a prompt known to trigger default images.
Participants showed particularly strong dissatisfaction with this image \ok{($Mean=2.4$, $SD=1.9$)}.
\ok{There was a significant difference in user satisfaction between Q3 (representing an exact match between prompt and image) and Q4 (representing a default image) with \ok{$OR=16.2$} and $p<0.05$.}

\subsection{Default Images on Midjourney}%
\label{sec:midjourneyresults}

\begin{figure*}[!thb]%
\centering%
    \includegraphics[width=.97\textwidth]{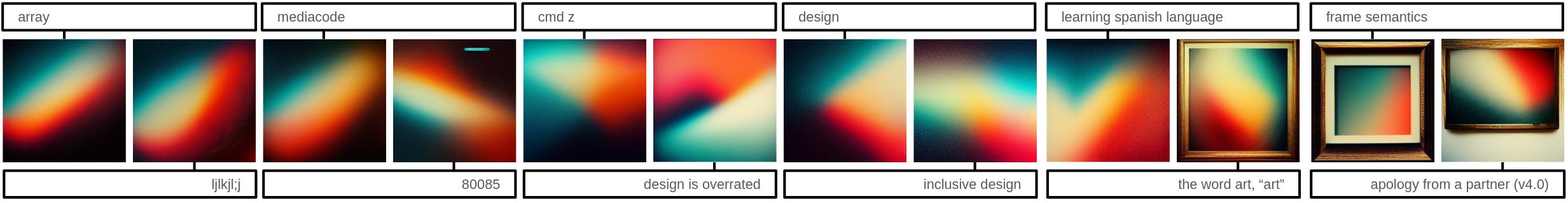}%
\\[.25\baselineskip]%
    \includegraphics[width=.97\textwidth]{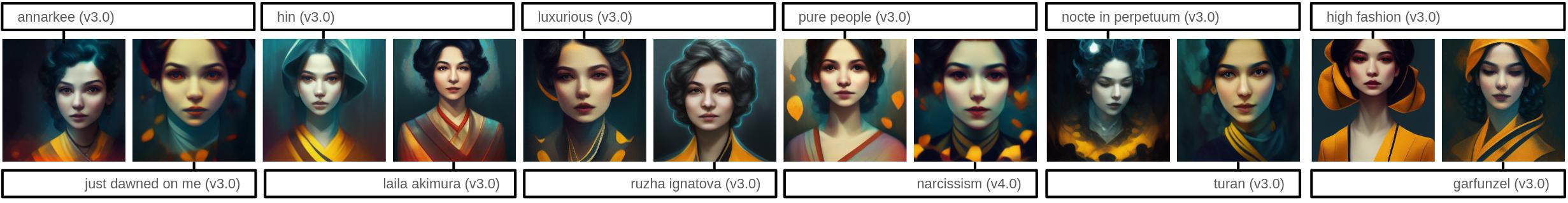}%
\\[.25\baselineskip]%
    \includegraphics[width=.97\textwidth]{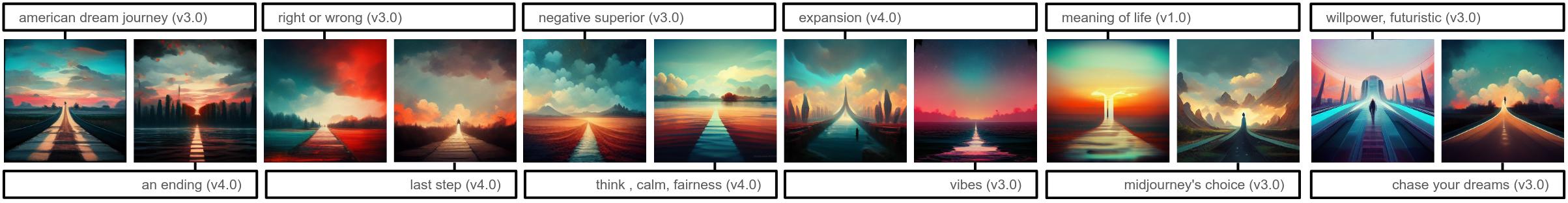}%
\\[.25\baselineskip]%
    \includegraphics[width=.97\textwidth]{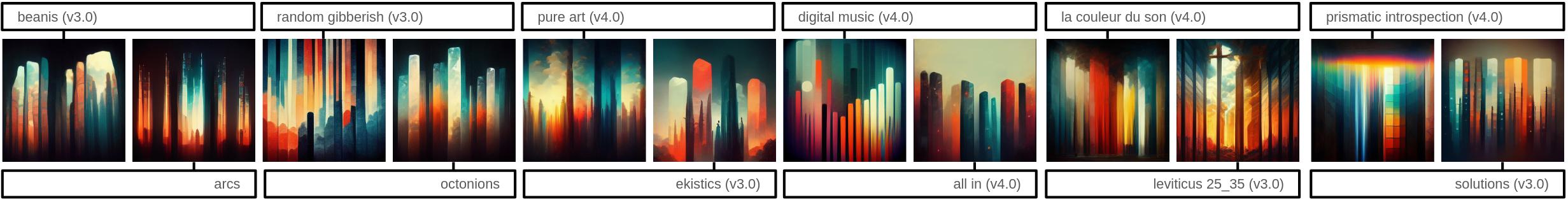}%
\\[.25\baselineskip]%
    \includegraphics[width=.97\textwidth]{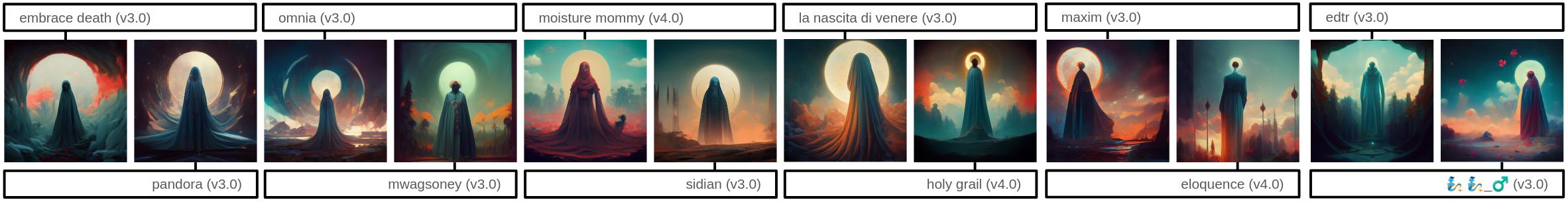}%
\\[.25\baselineskip]%
    \includegraphics[width=.97\textwidth]{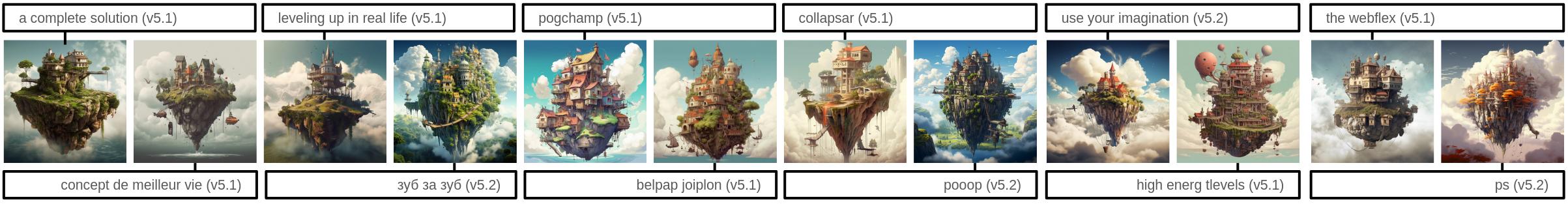}%
\\[.25\baselineskip]%
    \includegraphics[width=.97\textwidth]{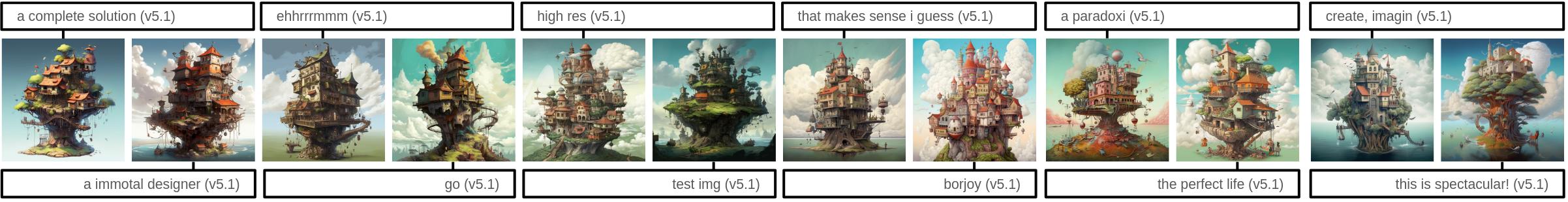}%
\\[.25\baselineskip]%
    \includegraphics[width=.97\textwidth]{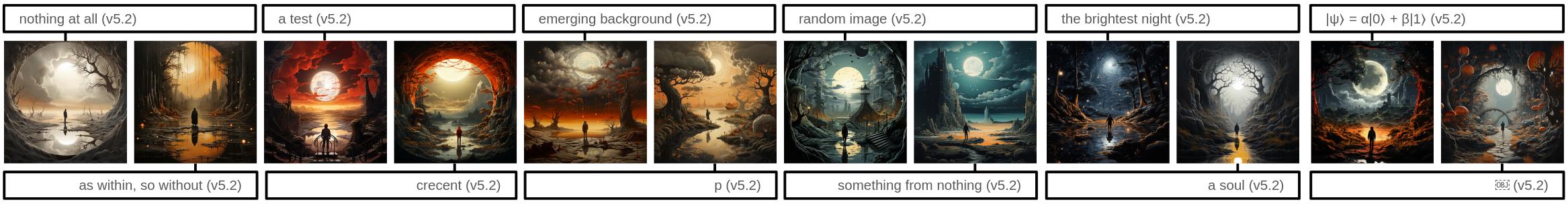}%
\caption{
\rev{%
Examples of default images and triggering prompts.
A curated set of default images and their associated prompts from our dataset, illustrating the diversity of motifs and styles that recur across unrelated inputs.
}%
}%
\Description{The figure depicts sets of default images from different Midjourney versions. The visual similarity in terms of subject, colors, style, and composition is uncanny.}%
\label{fig:analysis-results}%
\end{figure*}%

\subsubsection{Results}%
\label{sec:computationalfindings}

Our computational analysis of images on Midjourney resulted in the identification of
    \rev{4,715}
clusters of potential default images.
In total, the clusters comprise
    \rev{36,243}
images, representing
    \rev{4.84\%}
of our dataset.
This percentage refers only to our filtered dataset and parameters
    (i.e., less than 25~characters per prompt, lexical diversity $\geq$ 0.5, and semantic similarity \rev{$\leq$} 0.3)
and should not be interpreted as representative of the overall prevalence of default images on Midjourney.
    The mean number of images per cluster is \rev{7.69 (SD~=~9.04, Median~= 5, Min~= 3, Max~= 100)}.
    The lexical diversity of prompts within clusters ranges from \rev{0.5 to 1.0 (Mean~= 0.92, SD~=~0.12, Median~= 1.0)},
    and the semantic similarity from \rev{0.01 to 0.3 (Mean~= 0.2, SD~=~0.05, Median~= 0.2)}.
From smaller-scale experiments, we note the largest clusters were stable across multiple parameter settings.
%
%
Some selected canonical examples are depicted in
\autoref{fig:analysis-results} and 
Appendix~\ref{appendix:more-examples}.%

We make the following observations based on the results of the computational analysis:%
\begin{itemize}%
    \item We found evidence of short prompts, like the ones in our manual study, being in real-world use by Midjourney users.
    These prompts include, for instance, fantasy names, domain names, oxymorons, conceptual metaphors, abstract noun phrasing, poetic paradoxes and lyrical abstraction, surreal compound phrases, other neologisms and creative compounds, \rev{or typos}.
    This confirms that the six categories of prompts likely triggering default images in our manual study are realistic 
    and have practical relevance.
    \item  The clusters contain recurrent themes \rev{in the images,} identified in our earlier manual study, such as portrait-style images, intricate cityscapes, and dream-like imagery.
    This includes some of the iconic Midjourney motifs, 
    such as a person walking up an eternal staircase or road,
    a town on a floating island,
    abstract and diffuse color arrangements,
    and depictions of seemingly the same person.
    \item
    Major versions of Midjourney have different sets of default images (as expected), although some versions (such as version 3 and 4) share common default image motifs. This is \rev{likely} a result of training data being shared between training runs.    
    \item As models became more powerful, default images changed (cf. \autoref{fig:analysis-results} and Appendix~\ref{appendix:more-examples}).
    While in earlier models, default images tended to be more simplistic or abstract, 
    in recent systems default images are more complex and intricate.
    \item 
    There is more variation in the dataset than in our manual experiment. We did not encounter the default images from our manual study in our collected dataset. This is expected, as most images in our collected dataset were created with random seed values, as opposed to fixed seed values 
    in our manual experiments.
    \item 
    The variation makes detecting default images in real-world context more difficult.
    Motifs in default images seamlessly blend with each other based on (often random) seed values, further complicating real-world detection of default images in blackbox TTI models.%
\end{itemize}%
%
%
%
%
%
%
\section{Discussion}%
\label{sec:discussion}%
\subsection{Default Images in Text-to-Image Generation}%
Default images in TTI generation are an underexplored phenomenon.
We presented the first study on default images in Midjourney. 
When prompts are vague and contain unknown terms, Midjourney may generate default images, suggesting that the model defaults to familiar visual elements when it encounters words outside of its training data.

Our computational analysis of images on Midjourney reveals that default images are not a theoretical construct, they occur daily in response to real-world creative 
user prompts.
Some of the images we identified in our analysis represent iconic Midjourney failure modes that many users may have encountered, especially with the early Midjourney versions.
Our findings confirm that these default images are not generated randomly and, on the basis of our \rev{systematic} experiments, have a relatively narrow 
    thematic and formal range.
Uncanny elements emerged in common motifs that often depicted human portraits (usually women) and animals (especially birds), along with other botanical or organic forms.
Earlier versions of Midjourney have a different set of default images, often including motifs such as balloons, butterflies, fishes, floating islands, towns, 
\rev{or simply color pallettes and transitions}
(see \autoref{fig:teaser} and Appendix \ref{appendix:more-examples}).
Although many default images appear to share a common style, this straightforward association may obfuscate a more complex reality.
For instance, in model version 5, some default images appear related to surrealism and cartoonish grotesque 
(cf. \autoref{fig:teaser}), with
other identifiable references from art history, namely Dutch and Flemish Renaissance painting, particularly Hieronymus Bosch and Bruegel the Elder. 
AI-generated images are hybrid and composite, condensing multiple forms and references learned from the training dataset into a single visual output.
As \citeauthor{Wasielewski} points out ``style is a gradient'' \rev{\cite{Wasielewski}}. Therefore, an AI image is unlikely to correspond to a specific artistic movement, rather reflecting ``a fluctuating set of visual characteristics with highly fuzzy boundaries'' \cite{Wasielewski}.

Our findings also demonstrate that default images may negatively affect user satisfaction.
In the following section, we describe the characteristics of default images \rev{in eight postulates.}%
%
%
%
%
%
\subsection{
Characteristics of Default Images in Generative Models}%
%
We outline eight postulates that define key characteristics of default images, highlighting their occurrence, variability, and evolution as TTI models develop.%
\begin{itemize}%
    \item[P1.] \textit{Emergence of default images due to unknown inputs:}
        TTI models default to 
        certain visuals (``default images'') with distinct motifs 
        when they encounter inputs outside their training data.

    \item[P2.] \textit{Model-specific default images:}
        Each model version has its own set of default images, based on its training data.

    \item[P3.] \textit{Paradox of rarity and commonality:}
        Default images are both rare and common.
        Default images are rare in that they do not occur in response to most prompts written by Midjourney users.
        However, default images are also common in that they may occur in response to unknown inputs in a prompt.
        Default images blend seamlessly with other knappear to share a common style, this straightforward association may obfuscate a more complex reality. For instance, inown concepts in a prompt.

    \item[P4.] \textit{Influence of prompt specificity:}
        When users provide vague or ambiguous prompts, default images are more likely to emerge due to the model's reliance on generic visual patterns.

    \item[P5.] \textit{Paradox of creativity and default image frequency:}
        Default images are more common in `creative' prompts that seek to deviate and innovate.
    
    \item[P6.] \textit{Decreasing frequency with model advancement:}
        As the vocabulary of TTI models grows, default images become rarer.
        Therefore, newer models can be expected to produce fewer default images.
        As Midjourney releases new versions or integrates feedback, the nature and frequency of default images may change, reflecting shifts in the model’s learned visual patterns.

    \item[P7.]
        \textit{Modifying default images:}
        Styles, aesthetic modifiers, and compositional constraints can be applied to default images, altering their appearance while retaining their underlying motifs.
        We can adjust color palettes, lighting, perspective, or texture, yet traces of the default image’s motif and structure often remain detectable.

    \item[P8.] \textit{Variability of default images per prompt:}
        One prompt can produce several different default images. This is related to how Midjourney leverages stochastic sampling within its generative model to produce diverse visual outputs. 

\end{itemize}%
%
%
\subsection{Explainability and Alignment of Text-to-Image Generation Models}%
What does \textit{`absolute ownership'}, \textit{`nallopand'}, or a \textit{`protruding goofus'} look like?
Humans could, perhaps, imagine some creative imagery for these abstract concepts and lexical inventions.
However, machines struggle to produce a result (cf. \autoref{fig:teaser}).
This is not surprising, given that these terms were likely not part of the TTI models' training data.
While polysemia and ambiguity will likely continue to be an issue in the alignment of TTI models with human intent (consider, for instance, the polysemous prompt \textit{``a plant on a bank''}), default images are a different phenomenon -- a type of image that appears in response to prompts that are not understood by the model, revealing a misalignment between model and human intent.%

\citeauthor{daras2022} experimented with how gibberish prompts are understood by the model and suspected that TTI models have developed a `secret language' \cite{daras2022}.
Their hypothesis was that DALL-E 2 had made `hidden' associations between certain terms in text prompts and concepts in images, such as `\textit{Apoploe vesrreaitais}' meaning \textit{`bird'}.
\citeauthor{daras2022} found that these terms often produced consistent results in isolation but also sometimes in combination with other terms in prompts.
Our work is similar in that some input prompts reliably sample from the same region in latent space.
    We found that with prompts that include both known and unknown terms, the model often generates images from known terms, omitting unknown terms.
    However, when Midjourney does not ignore unknown terms, it may generate images sharing visual motifs with default images.
However, unlike in a `secret language', this is not a one-to-one mapping between prompt and resulting image. In our case, several unrelated prompts can produce variations of the same default image, as demonstrated in \rev{\autoref{fig:screenshot}.}

Because it is often unclear why the model produced a certain default image, the problem of default images touches on research in interpretability and Explainable AI (XAI).
Default images reflect inherent biases and patterns encoded in the TTI model.
In a sense, the default image is a nonce output and this output currently is the only mechanism of the TTI system to let the user know that the prompt is not understood.
    Other generative models, such as language models, include clear, explicit ways of letting users know that something is not understood or that they cannot respond (e.g., by refusing to answer for reasons of content policy).
    TTI models lack these mechanisms by default.
    In practice, language models internally handle such content and safety checks on Midjourney.
    Our prompts, however, pass these checks and obscure the fact that the model can not respond to the user prompt as expected.
    Instead of returning nothing, the TTI model generates default images.
    This could be termed a case of ostensive inferential communication \cite{sperber1995relevance,wilson2004relevance}, a model of communication where a speaker provides cues (ostension) to make their intended meaning clear, and the listener infers that meaning through a process of reasoning (inference).
This miscommunication is also a plausible explanation as to why user satisfaction is lower when the image is different from expectations.%
%
%
%
%
%
%
%
\subsection{The Future of Default Images}%
\label{sec:futuredefaultimages}%
%
We believe that in the future, the phenomenon of default images will transform for several reasons.
First, the available visual material, model vocabulary, and processing power will likely continue to increase.
As TTI models become more powerful \rev{in terms of training vocabulary}, they 
will likely generate fewer default images in response to user prompts.
Second, TTI models will continue to evolve. 
In particular, personalization could 
alter default image outputs in unpredictable ways, making it an important area for future research.
%
%
\rev{As personalization and context-awareness become standard, default images may shift from being shared phenomena to individualized experiences, tied to each user’s history, style preferences, or interaction patterns.
This shift could make systematic study more challenging, but it also opens questions about how visual defaults might adapt to, or even reinforce, a user's creative identity.}

However, the phenomenon of default images itself will likely not disappear entirely.
We will continue to see situations where words in prompts are not understood by the generative model, 
where users type incoherent prompts \cite{11084610} in error, or where users 
intentionally experiment with unfamiliar language, unusual phrasing, or invented terms to push the model beyond known outputs. In such cases, the model may revert to visual patterns that function as a ``fallback,'' even if these patterns become less recognizable as defaults in future \rev{generative} systems.

Finally, the interplay between user prompting practices and model architecture could shape new forms of defaults.
    For example, as prompt engineering becomes more sophisticated, communities might discover and share emergent default patterns, treating them as creative resources rather than artifacts of model limitations.
    Conversely, model designers may introduce mechanisms -- such as semantic interpolation or on-the-fly fine-tuning -- that actively mask or diversify defaults, making them harder to detect.
    The prompt rewriting employed in some commercial models under-the-hood already works towards this goal.%
%
%
%
\rev{%
\subsection{Design Recommendations for Text-to-Image Systems}%
In light of these observations, designers of text-to-image systems can draw several lessons from default images as a 
signal of model limitations.
First, default images point to a gap between user intent and model interpretation. Generative systems could surface this gap more transparently by clearly indicating which prompt elements are interpretable and which are not.
This could take the form of hints and gentle uncertainty cues in the user interface or interactive explanations that highlight the parts of a prompt that receive little or no semantic grounding in the generative model's internal representation.
Such feedback may support user awareness without interrupting creative flow.

Second, designers could introduce mechanisms that reduce the likelihood that default images become the only response to an unfamiliar prompt.
This could involve controlled fallback strategies that interpolate toward nearby known concepts while signaling that the system has done so,
or the introduction of explicit ``request for clarification'' prompts that preserve creative freedom while maintaining communicative clarity.
Some language models already ask for clarifications, and TTI generative models could be instruction-tuned to do the same.
These strategies could help prevent situations where users assume intentionality in outputs that originate from unrecognized terms.
On the other hand, many Midjourney users are likely not aware that some images in response to their prompts are default images.
For instance, the prompt ``modernika'' (see \autoref{fig:more-results} in Appendix \ref{appendix:more-examples}) could potentially depict a fantasy figure named Modernika. Yet, this image is a default image.
Whether or not the user needs to know the presence of this default image is a design choice.%

Third, personalization features raise questions about how default images evolve when generative systems adapt to individual users.
Designers may wish to consider whether personalized defaults should be transparent as defaults, or whether they should blend into the user's broader creative trajectory.
In other terms, designers will need to decide if users should be alerted of default images or not. 
Further, in creative contexts where exploration and play are central, personalized defaults could serve as material for experimentation, opening new creative pathways.
Conversely, in contexts that emphasize precision, such defaults may need more explicit framing so that users can distinguish between grounded outputs and fallback patterns.%

Fourth, as communities develop shared prompting practices, system designers could support ways of documenting, sharing, and surfacing these practices without reinforcing unwanted defaults.
Features that let users annotate prompts, sets of prompts, and discovered quirks could guide future users and inform model improvement.
Such community insight may also support internal design processes by revealing how defaults interact with the expectations and workflows of different creative user groups.

Finally, the presence of default images raises questions for future evaluation practices in TTI generation.
Designers could consider incorporating default-image behavior into model diagnostics, using default images as indicators or benchmarks of blind spots in the training corpus or the prompt parser.
Benchmarks that stress unknown terms or ambiguous phrasing may help identify where generative systems fail to align with user intent.
Such evaluation schemes would complement traditional performance metrics and help center, but also manage,  user experience in 
next-generation TTI models.%
}%

\transition{In the following we review some of the limitations of our study which can inspire \rev{additional} future work.}

\subsection{Limitations and Future Work}%
\label{sec:future-work}%
Our study offers a foundational investigation into the phenomenon of default images in TTI models. While the work is necessarily constrained in scope, these limitations highlight productive avenues for future research.

\rev{%
We presented a systematic empirical study without having access to the underlying training data or model.
It is our belief that default images arise from latent model behavior, likely due to gaps in the training data.
However, 
some system internals on Midjourney's platform (like the sampling mechanism or potential prompt rewriting) may contribute to the phenomenon of default images.
For instance, the many-to-many relationship between prompts and images observed in \autoref{fig:screenshot} can be attributed to diversity mechanisms in Midjourney’s stochastic sampling. 
On the other hand, other mechanisms, such as the LLM-based moderation feature, can likely be excluded as an alternative explanation for default images.
Moderation either lets prompts pass or not, and moderation is therefore not mechanistically involved in generating default images.
Nevertheless, important interpretative 
limitations around causality remain due to the opaque nature of Midjourney's proprietary platform.
Without knowing whether and how the system is intervening in its inputs and outputs, it is difficult to fully interpret 
whether default images reflect model behavior or platform design decisions.%
}%

Our focus on Midjourney limits the external validity of findings to other TTI systems. We selected this scope to allow in-depth qualitative and quantitative analysis within a consistent model environment.
Future research could adopt a comparative lens, investigating default images in other TTI generation systems, such as Stable Diffusion \cite{Rombach_2022_CVPR}, Imagen~\cite{2205.11487.pdf}, Flux~\cite{labs2025flux1kontextflowmatching}, Janus~\cite{janus}, \rev{or 
video generation models,} potentially revealing differences in model capabilities and biases beyond canonical benchmarks.

Due to the vastness of Midjourney’s latent space, our analysis cannot be exhaustive \cite{2507.17922.pdf,3491102.3501825.pdf}.
Each model version introduces its own set of default images, with some occurring far more frequently than others (cf. \autoref{fig:frequencies}). As a result, our work can only provide a partial view of the phenomenon.
Nonetheless, certain defaults (like the Lady-Birdhead image), are common and straightforward to identify even in small-scale manual studies, suggesting that foundational insights can be gained without full coverage.
Future work could extend our approach to systematically map default images across different model versions.
Default images 
    could form the basis of a benchmark for model evaluation, capturing how systems respond to unknown or ambiguous inputs.
    Such a benchmark would complement existing technical metrics by foregrounding model behavior in edge-case scenarios encountered in creative practice.


We restricted our analysis to short prompts to maximize the likelihood of encountering default images while keeping computational demands manageable.
This constraint limits conclusions about default images that may emerge in more elaborate prompt contexts.
Future studies could explore how prompt length, structure, modifiers, and semantic specificity influence the emergence and variability of default images.
From a preliminary study, we saw different images arise when combining large amounts of our prompts. Future work could study whether these prompts are default images, or whether there is a change in model behavior when prompted with longer prompts.
Negative prompts, weighted prompts, and image prompts were also not in scope of our foundational study and left to future work.
Further, our sensitivity analysis and ablation studies were limited and offer new ground for research.
Future work could systematically examine the role of seeds, negative prompts, and other parameters  
in shaping default outputs.

Calculating the full similarity matrix for images becomes computationally prohibitive for large numbers of images.
The memory requirement was substantial for our sample, but still feasible.
Instead of the full similarity matrix, future work could
employ approximate nearest neighbor search and clustering methods, e.g.
    by computing a sparse similarity matrix and retrieving only the top-k most similar images (e.g., top-100 images).

We did not directly investigate bias in default images in this work, yet their occurrence in the absence of explicit text guidance potentially provides an opportunity to study ``raw'' biases in TTI models.
Building on related studies on bias in TTI models (e.g., \cite{Vasilev2024S137,Elsharif2025122636,Saravanan202384,Kandwal202473,luccioni2023stablebiasanalyzingsocietal,10.1145/3706599.3719678,jha2024visageglobalscaleanalysisvisual,vice2023quantifyingbiastexttoimagegenerative}), future research could analyze demographic, cultural, or aesthetic biases embedded in default images.

Studying default images from a color theory perspective could also provide new insights on prevailing harmonies and contrasts and contribute to an understanding of the symbolism and potential impact of such associations.
Exploring how default images tie into the creative practice could be an interesting avenue for future studies.
Future work could examine the implications of default images for creativity support, for example through Construal Level Theory~\cite{Trope2010-TROCTO} or other psychological frameworks.
%

Finally, while our study focused on characterizing default images, mechanisms for detecting and handling them on-the-fly remain an open challenge.
Future work could explore real-time detection pipelines, interface designs that highlight misunderstood prompt segments, and techniques to mitigate unintended defaults without reducing creative potential.
Such approaches could improve prompt engineering practices and enhance user satisfaction.

\label{sec:limitations}%
\section{Conclusion}%
We presented the first systematic investigation of default images in Midjourney. Default images are images with a certain motif 
    that occur in response to a variety of dissimilar prompts.
Through an initial study with manual prompt‐generation and affinity diagramming, we identified ten canonical default images.
Scaling up to a larger set of images, we presented a simple method for identifying default images in Midjourney.
Through a large-scale computational analysis of Midjourney images, we demonstrated that default images are not just a theoretical construct, but occur in response to real-world user prompts.
Our user study showed that default images may negatively influence user satisfaction.
Based on our findings, we distilled eight postulates characterizing the
properties of default images.
By defining the notion of default images, our work lays the groundwork for future research on the limits of generative 
    systems.%

\begin{acks}
This work is supported in part by the Natural Sciences and Engineering Research Council of Canada (NSERC) Vanier Scholarship CGV-186947.
The last author gratefully acknowledges support from the University of Oulu's
Centre for Applied Computing (CAC) and
the NOSTE development programme for education.
\end{acks}

\bibliographystyle{ACM-Reference-Format}
\bibliography{paper}

@inproceedings{Promptify,
author = {Mahdavi Goloujeh, Atefeh and Sullivan, Anne and Magerko, Brian},
title = {Is It AI or Is It Me? Understanding Users’ Prompt Journey with Text-to-Image Generative AI Tools},
year = {2024},
isbn = {9798400703300},
publisher = {Association for Computing Machinery},
address = {New York, NY, USA},
doi = {10.1145/3613904.3642861},
booktitle = {Proceedings of the 2024 CHI Conference on Human Factors in Computing Systems},
articleno = {183},
numpages = {13},
series = {CHI '24}
}

@inproceedings{3613904.3642861.pdf,
author = {Mahdavi Goloujeh, Atefeh and Sullivan, Anne and Magerko, Brian},
title = {Is It AI or Is It Me? Understanding Users’ Prompt Journey with Text-to-Image Generative AI Tools},
year = {2024},
isbn = {9798400703300},
publisher = {Association for Computing Machinery},
address = {New York, NY, USA},
doi = {10.1145/3613904.3642861},
booktitle = {Proceedings of the 2024 CHI Conference on Human Factors in Computing Systems},
articleno = {183},
numpages = {13},
series = {CHI '24}
}

@inproceedings{2403.12075.pdf,
author = {Quaye, Jessica and Parrish, Alicia and Inel, Oana and Rastogi, Charvi and Kirk, Hannah Rose and Kahng, Minsuk and Van Liemt, Erin and Bartolo, Max and Tsang, Jess and White, Justin and Clement, Nathan and Mosquera, Rafael and Ciro, Juan and Janapa Reddi, Vijay and Aroyo, Lora},
title = {Adversarial Nibbler: An Open Red-Teaming Method for Identifying Diverse Harms in Text-to-Image Generation},
year = {2024},
isbn = {9798400704505},
publisher = {Association for Computing Machinery},
address = {New York, NY, USA},
doi = {10.1145/3630106.3658913},
booktitle = {Proceedings of the 2024 ACM Conference on Fairness, Accountability, and Transparency},
pages = {388–406},
numpages = {19},
series = {FAccT '24}
}

@inproceedings{3613905.3650947.pdf,
author = {Mahdavi Goloujeh, Atefeh and Sullivan, Anne and Magerko, Brian},
title = {The Social Construction of Generative AI Prompts},
year = {2024},
isbn = {9798400703317},
publisher = {Association for Computing Machinery},
address = {New York, NY, USA},
doi = {10.1145/3613905.3650947},
booktitle = {Extended Abstracts of the CHI Conference on Human Factors in Computing Systems},
articleno = {320},
numpages = {7},
series = {CHI EA '24}
}

@inproceedings{3613904.3642016.pdf,
author = {Arawjo, Ian and Swoopes, Chelse and Vaithilingam, Priyan and Wattenberg, Martin and Glassman, Elena L.},
title = {ChainForge: A Visual Toolkit for Prompt Engineering and LLM Hypothesis Testing},
year = {2024},
isbn = {9798400703300},
publisher = {Association for Computing Machinery},
address = {New York, NY, USA},
doi = {10.1145/3613904.3642016},
booktitle = {Proceedings of the 2024 CHI Conference on Human Factors in Computing Systems},
articleno = {304},
numpages = {18},
series = {CHI '24}
}

@inproceedings{3706598.3713166.pdf,
author = {Subramonyam, Hari and Thakkar, Divy and Ku, Andrew and Dieber, Juergen and Sinha, Anoop K.},
title = {Prototyping with Prompts: Emerging Approaches and Challenges in Generative AI Design for Collaborative Software Teams},
year = {2025},
isbn = {9798400713941},
publisher = {Association for Computing Machinery},
address = {New York, NY, USA},
doi = {10.1145/3706598.3713166},
booktitle = {Proceedings of the 2025 CHI Conference on Human Factors in Computing Systems},
articleno = {882},
numpages = {22},
series = {CHI '25}
}

@ARTICLE{Riveiro2021,
	author = {Riveiro, Maria and Thill, Serge},
	title = {“That's (not) the output I expected!” On the role of end user expectations in creating explanations of AI systems},
	year = {2021},
	journal = {Artificial Intelligence},
	volume = {298},
	doi = {10.1016/j.artint.2021.103507},
}

@CONFERENCE{Louie2020,
	author = {Louie, Ryan and Coenen, Andy and Huang, Cheng Zhi and Terry, Michael and Cai, Carrie J.},
	title = {Novice-AI Music Co-Creation via AI-Steering Tools for Deep Generative Models},
	year = {2020},
	journal = {Conference on Human Factors in Computing Systems - Proceedings},
	doi = {10.1145/3313831.3376739},
}

@CONFERENCE{Honeycutt202063,
	author = {Honeycutt, Donald R. and Nourani, Mahsan and Ragan, Eric D.},
	title = {Soliciting Human-in-the-Loop User Feedback for Interactive Machine Learning Reduces User Trust and Impressions of Model Accuracy},
	year = {2020},
	journal = {Proceedings of the AAAI Conference on Human Computation and Crowdsourcing},
	volume = {8},
	pages = {63 – 72},
	doi = {10.1609/hcomp.v8i1.7464},
}

@ARTICLE{Kreminski2025,
	author = {Kreminski, Max},
	title = {Endless forms most similar: The death of the author in AI-supported art},
	year = {2025},
	journal = {AI and Society},
	doi = {10.1007/s00146-025-02326-6},
}

@inproceedings{ICCC-2023_paper_104.pdf,
  author       = {Charlotte Bird},
  editor       = {Alison Pease and
                  Jo{\~{a}}o Miguel Cunha and
                  Maya Ackerman and
                  Daniel G. Brown},
  title        = {Evaluating Prompt Engineering as a Creative Practice},
  booktitle    = {Proceedings of the 14th International Conference on Computational
                  Creativity, {ICCC} 2023, Ontario, Canada, June 19-23, 2023},
  pages        = {352--356},
  publisher    = {Association for Computational Creativity {(ACC)}},
  year         = {2023},
  url          = {https://computationalcreativity.net/iccc23/papers/ICCC-2023\_paper\_104.pdf},
}

@misc{white2022schrodingersbatdiffusionmodels,
      title={Schr\"{o}dinger's Bat: Diffusion Models Sometimes Generate Polysemous Words in Superposition}, 
      author={Jennifer C. White and Ryan Cotterell},
      year={2022},
      eprint={2211.13095},
      archivePrefix={arXiv},
}

@inproceedings{maldaner2025miragemultimodelinterfacereviewing,
      title={MIRAGE: Multi-model Interface for Reviewing and Auditing Generative Text-to-Image AI}, 
      author={Matheus Kunzler Maldaner and Wesley Hanwen Deng and Jason Hong and Ken Holstein and Motahhare Eslami},
      year={2025},
booktitle={Proceedings of the Twelfth AAAI Conference on Human Computation and Crowdsourcing},
      url={https://www.humancomputation.com/assets/wip_2024/HCOMP_24_WIP_4.pdf},
}

@inproceedings{2203.06026,
title={The Role of ImageNet Classes in Fr\'echet Inception Distance},
author={Tuomas Kynk{\"a}{\"a}nniemi and Tero Karras and Miika Aittala and Timo Aila and Jaakko Lehtinen},
booktitle={The Eleventh International Conference on Learning Representations },
year={2023},
url={https://openreview.net/forum?id=4oXTQ6m_ws8}
}

@article{Morris,
author = {Morris, Meredith Ringel},
title = {Prompting Considered Harmful},
year = {2024},
issue_date = {December 2024},
publisher = {Association for Computing Machinery},
address = {New York, NY, USA},
volume = {67},
number = {12},
issn = {0001-0782},
url = {https://doi.org/10.1145/3673861},
doi = {10.1145/3673861},
journal = {Commun. ACM},
month = nov,
pages = {28–30},
numpages = {3}
}

@book{Wasielewski,
    author = {Wasielewski, Amanda},
    title = {Computational Formalism: Art History and Machine Learning},
    publisher = {The MIT Press},
    year = {2023},
    month = {05},
    isbn = {9780262374736},
    doi = {10.7551/mitpress/14268.001.0001},
}

@book{ALLOA,
author={Emmanuel Alloa},
year={2021},
title={Looking Through Images. A Phenomenology of Visual Media},
publisher={Columbia University Press},
address={New York, NY},
}

@article{PRADA,
author={Martín Prada, Juan},
title={AI-based generative image production systems in the artistic problematisation of the past: the thematisation of memory and temporality in ``AI art''},
year={2025},
journal={AI \& Society},
number={5},
volume={40},
pages={3271--3282},
doi={10.1007/s00146-024-02163-z},
}

@article{OFFERT,
author={Offert, Fabian and Bell, Peter},
title={Perceptual bias and technical metapictures: Critical machine vision as a humanities challenge},
year={2021},
journal={AI \& Society},
number={4},
volume={36},
pages={1133--1144},
doi={10.1007/s00146-020-01058-z},
}

@incollection{palgrave,
  title = {The Cultivated Practices of Text-to-Image Generation},
  author = {Jonas Oppenlaender},
  year = {2024},
  publisher = {Palgrave Macmillan},
  isbn = {978-3-031-66527-1},
  booktitle = {Humane Autonomous Technology. Re-thinking Experience with and in Intelligent Systems},
  doi = {10.1007/978-3-031-66528-8_14},
}

@article{prompting-ai-art,
  author = {Jonas Oppenlaender and Rhema Linder and Johanna Silvennoinen},
  title = {Prompting {AI} Art: An Investigation into the Creative Skill of Prompt Engineering},
  year = {2024},
  journal = {International Journal of Human–Computer Interaction},
  series = {IJHCI},
  volume = {},
  number = {},
  pages = {1--23},
  publisher = {Taylor \& Francis},
  doi = {10.1080/10447318.2024.2431761},
}

@misc{2407.17493.pdf,
      title={Model Collapse in the Self-Consuming Chain of Diffusion Finetuning: A Novel Perspective from Quantitative Trait Modeling}, 
      author={Youngseok Yoon and Dainong Hu and Iain Weissburg and Yao Qin and Haewon Jeong},
      year={2025},
      eprint={2407.17493},
      archivePrefix={arXiv},
}

@article{Runco01012012,
author = {Mark A. Runco and Garrett J. Jaeger},
title = {The Standard Definition of Creativity},
journal = {Creativity Research Journal},
volume = {24},
number = {1},
pages = {92--96},
year = {2012},
publisher = {Routledge},
doi = {10.1080/10400419.2012.650092},
}

@ARTICLE{Zhang2024,
	author = {Zhang, Guoqing and Zhou, Ruixin and Zheng, Yuhui and Li, Baozhu},
	title = {Binary Noise Guidance Learning for Remote Sensing Image-to-Image Translation},
	year = {2024},
	journal = {Remote Sensing},
	volume = {16},
	number = {1},
	doi = {10.3390/rs16010065},
}

@ARTICLE{Zhang20217789,
	author = {Zhang, Min and Li, Chunye and Zhou, Zhiping},
	title = {Text to image synthesis using multi-generator text conditioned generative adversarial networks},
	year = {2021},
	journal = {Multimedia Tools and Applications},
	volume = {80},
	number = {5},
	pages = {7789–7803},
	doi = {10.1007/s11042-020-09965-5},
}

@INPROCEEDINGS{10315510,
  author={Tominaga, Rihito and Seo, Masataka},
  booktitle={2023 IEEE 12th Global Conference on Consumer Electronics (GCCE)}, 
  title={Novel Regularization Method for Text-to-Image Generation Using Deep Learning}, 
  year={2023},
  volume={},
  number={},
  pages={842-844},
  doi={10.1109/GCCE59613.2023.10315510},
}

@CONFERENCE{Liu20212081,
	author = {Liu, Hongbin and Jia, Jinyuan and Qu, Wenjie and Gong, Neil Zhenqiang},
	title = {EncoderMI: Membership Inference against Pre-trained Encoders in Contrastive Learning},
	year = {2021},
	journal = {Proceedings of the ACM Conference on Computer and Communications Security},
	pages = {2081 – 2095},
	doi = {10.1145/3460120.3484749},
}

@incollection{MANOVICH,
title={From Representation to Prediction: Theorizing the AI Image},
author={Lev Manovich},
year={2024},
editor={Lev Manovich and Emanuele Arielli},
booktitle={Artificial Aesthetics: Generative AI, Art and Visual Media},
pages={75-99},
}

@inproceedings{2508.06065.pdf,
author={Daniel Lee and Nikhil Sharma and Donghoon Shin and DaEun Choi and Harsh Sharma and Jeonghwan Kim and Heng Ji},
title={ThematicPlane: Bridging Tacit User Intent and Latent Spaces for Image Generation},
booktitle={UIST Adjunct '25: Adjunct Proceedings of the 38th Annual ACM Symposium on User Interface Software and Technology},
year={2025},
doi={10.1145/3746058.3758376},
}

@CONFERENCE{Jiang202590485,
	author = {Jiang, Yue and Lin, Haokun and Bai, Yang and Peng, Bo and Liu, Zhili and Lyu, Yueming and Yang, Yong and Zheng, Xing and Dong, Jing},
	title = {Image-level Memorization Detection Via Inversion-based Inference Perurbation},
	year = {2025},
	journal = {13th International Conference on Learning Representations, ICLR 2025},
	pages = {90485 – 90504},
}

@ARTICLE{Wu2021,
	author = {Wu, Zhenyu and Wang, Zhaowen and Yuan, Ye and Zhang, Jianming and Wang, Zhangyang and Jin, Hailin},
	title = {Black-box diagnosis and calibration on gan intra-mode collapse: A pilot study},
	year = {2021},
	journal = {ACM Transactions on Multimedia Computing, Communications and Applications},
	volume = {17},
	number = {3s},
	doi = {10.1145/3472768},
}

@ARTICLE{Gong2023,
	author = {Gong, Yanxiang and Zhong, Minjiang and Ji, Yang and Xie, Mei and Ma, Xin},
	title = {Distribution constraining for combating mode collapse in generative adversarial networks},
	year = {2023},
	journal = {Journal of Electronic Imaging},
	volume = {32},
	number = {4},
	doi = {10.1117/1.JEI.32.4.043029},
}

@misc{mishra2025promptaidpromptexplorationperturbation,
      title={PromptAid: Prompt Exploration, Perturbation, Testing and Iteration using Visual Analytics for Large Language Models}, 
      author={Aditi Mishra and Utkarsh Soni and Anjana Arunkumar and Jinbin Huang and Bum Chul Kwon and Chris Bryan},
      year={2025},
      eprint={2304.01964},
      archivePrefix={arXiv},
      url={https://arxiv.org/abs/2304.01964}, 
}

@article{Text-35672-1-2-20241013.pdf, title={Responsible Crowdsourcing for Responsible Generative AI: Engaging Crowds in AI Auditing and Evaluation}, volume={12}, url={https://ojs.aaai.org/index.php/HCOMP/article/view/31609}, DOI={10.1609/hcomp.v12i1.31609}, abstractNote={With the rise of generative AI (GenAI), there has been an increased need for participation by large and diverse user bases in AI evaluation and auditing. GenAI developers are increasingly adopting crowdsourcing approaches to test and audit their AI products and services. However, it remains an open question how to design and deploy responsible and effective crowdsourcing pipelines for AI auditing and evaluation. This workshop aims to take a step towards bridging this gap. Our interdisciplinary team of organizers will work with workshop participants to explore several key questions, such as how to improve the output quality and workers’ productivity for GenAI evaluation crowdsourcing tasks compared to discriminative AI systems, how to guide crowds in auditing problematic AI-generated content while managing their psychological impact, ensuring marginalized voices are heard, and setting up responsible and effective crowdsourcing pipelines for real-world GenAI evaluation. We hope this workshop will produce a research agenda and best practices for designing responsible crowd-based approaches to AI auditing and evaluation.}, number={1}, journal={Proceedings of the AAAI Conference on Human Computation and Crowdsourcing}, author={Deng, Wesley Hanwen and Yurrita, Mireia and Díaz, Mark and Suh, Jina and Judd, Nick and Groves, Lara and Shen, Hong and Eslami, Motahhare and Holstein, Kenneth}, year={2024}, month={Oct.}, pages={148-150} }

@article{Choi_Kim_Song_2022, title={Style-Guided and Disentangled Representation for Robust Image-to-Image Translation}, volume={36}, url={https://ojs.aaai.org/index.php/AAAI/article/view/19924}, DOI={10.1609/aaai.v36i1.19924},
number={1}, journal={Proceedings of the AAAI Conference on Artificial Intelligence}, author={Choi, Jaewoong and Kim, Daeha and Song, Byung Cheol}, year={2022}, month={Jun.}, pages={463-471} }

@inproceedings{GenAssist,
author = {Huh, Mina and Peng, Yi-Hao and Pavel, Amy},
title = {GenAssist: Making Image Generation Accessible},
year = {2023},
isbn = {9798400701320},
publisher = {Association for Computing Machinery},
address = {New York, NY, USA},
doi = {10.1145/3586183.3606735},
booktitle = {Proceedings of the 36th Annual ACM Symposium on User Interface Software and Technology},
articleno = {38},
numpages = {17},
series = {UIST '23}
}

@INPROCEEDINGS{10334839,
  author={Hashemi, Seyyed Morteza and Aliniya, Parvaneh and Razzaghi, Parvin},
  booktitle={2023 31st International Conference on Electrical Engineering (ICEE)}, 
  title={Connective Reconstruction-Based Novelty Detection}, 
  year={2023},
  volume={},
  number={},
  pages={563-569},
  doi={10.1109/ICEE59167.2023.10334839},
}

@article{s40747-022-00924-1.pdf,
  author    = {Yao Gou and Min Li and Yu Song and Yujie He and Litao Wang},
  title     = {Multi-feature contrastive learning for unpaired image-to-image translation},
  journal   = {Complex \& Intelligent Systems},
  year      = {2023},
  volume    = {9},
  number    = {4},
  pages     = {4111--4122},
  doi       = {10.1007/s40747-022-00924-1},
  issn      = {2198-6053},
}

@InProceedings{978-3-031-90341-0_20,
author="Barsha, Farhat Lamia
and Eberle, William",
editor="Arabnia, Hamid R.
and Deligiannidis, Leonidas
and Ghareh Mohammadi, Farid
and Amirian, Soheyla
and Shenavarmasouleh, Farzan",
title="Early Detection of Mode Collapse in GANs Through Loss Monitoring",
booktitle="Computational Science and Computational Intelligence",
year="2025",
publisher="Springer Nature Switzerland",
address="Cham",
pages="268--283",
isbn="978-3-031-90341-0"
}

@misc{1910.11626.pdf,
      title={Seeing What a GAN Cannot Generate}, 
      author={David Bau and Jun-Yan Zhu and Jonas Wulff and William Peebles and Hendrik Strobelt and Bolei Zhou and Antonio Torralba},
      year={2019},
      eprint={1910.11626},
      archivePrefix={arXiv},
}

@incollection{ARIELLI,
author={Emanuele Arielli},
year={2024},
title={Made By and For Humans? The Issue of Aesthetic Alignment},
editor={Lev Manovich and Emanuele Arielli},
booktitle={Artificial Aesthetics: Generative AI, Art and Visual Media},
pages={180-193},
}

@InProceedings{Duym_2025_WACV,
    author    = {Duym, Jens and Mogrovejo, Jos\'e Antonio Oramas and Anwar, Ali},
    title     = {Quantifying Generative Stability: Mode Collapse Entropy Score for Mode Diversity Evaluation},
    booktitle = {Proceedings of the Winter Conference on Applications of Computer Vision (WACV) Workshops},
    month     = {February},
    year      = {2025},
    pages     = {187-196},
url={https://openaccess.thecvf.com/content/WACV2025W/ImageQuality/html/Duym_Quantifying_Generative_Stability_Mode_Collapse_Entropy_Score_for_Mode_Diversity_WACVW_2025_paper.html},
}

@misc{2505.08803.pdf,
      title={Multi-modal Synthetic Data Training and Model Collapse: Insights from VLMs and Diffusion Models}, 
      author={Zizhao Hu and Mohammad Rostami and Jesse Thomason},
      year={2025},
      eprint={2505.08803},
      archivePrefix={arXiv},
}

@inproceedings{NIPS-2014-generative-adversarial-nets-Paper.pdf,
 author = {Goodfellow, Ian J. and Pouget-Abadie, Jean and Mirza, Mehdi and Xu, Bing and Warde-Farley, David and Ozair, Sherjil and Courville, Aaron and Bengio, Yoshua},
 booktitle = {Advances in Neural Information Processing Systems},
 editor = {Z. Ghahramani and M. Welling and C. Cortes and N. Lawrence and K.Q. Weinberger},
 pages = {},
 publisher = {Curran Associates, Inc.},
 title = {Generative Adversarial Nets},
 url = {https://proceedings.neurips.cc/paper_files/paper/2014/file/f033ed80deb0234979a61f95710dbe25-Paper.pdf},
 volume = {27},
 year = {2014}
}

@inproceedings{CLIPScore,
    title = "{CLIPS}core: A Reference-free Evaluation Metric for Image Captioning",
    author = "Hessel, Jack  and
      Holtzman, Ari  and
      Forbes, Maxwell  and
      Le Bras, Ronan  and
      Choi, Yejin",
    editor = "Moens, Marie-Francine  and
      Huang, Xuanjing  and
      Specia, Lucia  and
      Yih, Scott Wen-tau",
    booktitle = "Proceedings of the 2021 Conference on Empirical Methods in Natural Language Processing",
    month = nov,
    year = "2021",
    address = "Online and Punta Cana, Dominican Republic",
    publisher = "Association for Computational Linguistics",
    url = "https://aclanthology.org/2021.emnlp-main.595/",
    doi = "10.18653/v1/2021.emnlp-main.595",
    pages = "7514--7528",
}

@inproceedings{VQAScore,
author = {Lin, Zhiqiu and Pathak, Deepak and Li, Baiqi and Li, Jiayao and Xia, Xide and Neubig, Graham and Zhang, Pengchuan and Ramanan, Deva},
title = {Evaluating Text-to-Visual Generation with Image-to-Text Generation},
year = {2024},
isbn = {978-3-031-72672-9},
publisher = {Springer-Verlag},
address = {Berlin, Heidelberg},
url = {https://doi.org/10.1007/978-3-031-72673-6_20},
doi = {10.1007/978-3-031-72673-6_20},
booktitle = {Computer Vision – ECCV 2024: 18th European Conference, Milan, Italy, September 29–October 4, 2024, Proceedings, Part IX},
pages = {366–384},
numpages = {19},
}

@inproceedings{3491102.3501825.pdf,
author = {Liu, Vivian and Chilton, Lydia B},
title = {Design Guidelines for Prompt Engineering Text-to-Image Generative Models},
year = {2022},
isbn = {9781450391573},
publisher = {Association for Computing Machinery},
address = {New York, NY, USA},
doi = {10.1145/3491102.3501825},
booktitle = {Proceedings of the 2022 CHI Conference on Human Factors in Computing Systems},
articleno = {384},
numpages = {23},
series = {CHI '22}
}

@misc{pagpag,
url={https://en.wikipedia.org/wiki/Pagpag},
author={{Wikiedpia editors}},
year={n.d.},
publisher={Wikiedpia},
}

@misc{ghibli,
url={https://www.forbes.com/sites/danidiplacido/2025/03/27/the-ai-generated-studio-ghibli-trend-explained/},
title={The AI-Generated Studio Ghibli Trend, Explained},
author={Dani Di Placido},
year={2025},
publisher={Forbes},
}

@inproceedings{3527927.3532792.pdf,
author = {Qiao, Han and Liu, Vivian and Chilton, Lydia},
title = {Initial Images: Using Image Prompts to Improve Subject Representation in Multimodal AI Generated Art},
year = {2022},
isbn = {9781450393270},
publisher = {Association for Computing Machinery},
address = {New York, NY, USA},
doi = {10.1145/3527927.3532792},
booktitle = {Proceedings of the 14th Conference on Creativity and Cognition},
pages = {15–28},
numpages = {14},
series = {C\&C '22}
}

@article{taxonomy,
  author = {Jonas Oppenlaender},
  title = {A Taxonomy of Prompt Modifiers for Text-To-Image Generation},
  year = {2023},
  journal = {Behaviour \& Information Technology},
  publisher = {Taylor \& Francis},
  doi = {10.1080/0144929X.2023.2286532},
  pages = {3763–3776},
  volume={43},
  number={15},
}

@inproceedings{creativity,
  author = {Jonas Oppenlaender},
  title = {The Creativity of Text-to-Image Generation},
  year = {2022},
  isbn = {9781450399555},
  publisher = {Association for Computing Machinery},
  address = {New York, NY, USA},
  doi = {10.1145/3569219.3569352},
  booktitle = {25th International Academic Mindtrek Conference},
  pages = {192–202},
  numpages = {11},
  series = {Academic Mindtrek 2022},
}

@INPROCEEDINGS{1801.03924.pdf,
author = {Zhang, Richard and Isola, Phillip and Efros, Alexei A. and Shechtman, Eli and Wang, Oliver },
booktitle = { 2018 IEEE/CVF Conference on Computer Vision and Pattern Recognition (CVPR) },
title = {{ The Unreasonable Effectiveness of Deep Features as a Perceptual Metric }},
year = {2018},
volume = {},
ISSN = {},
pages = {586-595},
doi = {10.1109/CVPR.2018.00068},
publisher = {IEEE Computer Society},
address = {Los Alamitos, CA, USA},
}

@ARTICLE{1284396,
  author={Jalobeanu, A. and Blanc-Feraud, L. and Zerubia, J.},
  journal={IEEE Transactions on Image Processing}, 
  title={An adaptive Gaussian model for satellite image deblurring}, 
  year={2004},
  volume={13},
  number={4},
  pages={613-621},
  doi={10.1109/TIP.2003.819969}
}

@article{10.1002/widm.53,
author = {Murtagh, Fionn and Contreras, Pedro},
title = {Algorithms for hierarchical clustering: an overview},
journal = {WIREs Data Mining and Knowledge Discovery},
volume = {2},
number = {1},
pages = {86-97},
doi = {https://doi.org/10.1002/widm.53},
year = {2012},
}

@ARTICLE{1284395,
  author={Zhou Wang and Bovik, A.C. and Sheikh, H.R. and Simoncelli, E.P.},
  journal={IEEE Transactions on Image Processing}, 
  title={Image quality assessment: from error visibility to structural similarity}, 
  year={2004},
  volume={13},
  number={4},
  pages={600-612},
  doi={10.1109/TIP.2003.819861},
}

@InProceedings{gafni:2022,
author="Gafni, Oran
and Polyak, Adam
and Ashual, Oron
and Sheynin, Shelly
and Parikh, Devi
and Taigman, Yaniv",
editor="Avidan, Shai
and Brostow, Gabriel
and Ciss{\'e}, Moustapha
and Farinella, Giovanni Maria
and Hassner, Tal",
title="Make-A-Scene: Scene-Based Text-to-Image Generation with Human Priors",
booktitle="Computer Vision -- ECCV 2022",
year="2022",
publisher="Springer Nature Switzerland",
address="Cham",
pages="89--106",
isbn="978-3-031-19784-0"
}

@misc{SolidGoldMagikarp,
title={SolidGoldMagikarp (plus, prompt generation)},
author={Rumbelow, Jessica and Watkins, M.},
year={2023},
url={https://www.lesswrong.com/posts/aPeJE8bSo6rAFoLqg/solidgoldmagikarp-plus-prompt-generation},
publisher={lesswrong.com},
}

@misc{SolidGoldMagikarp2,
title={SolidGoldMagikarp II: Technical details and more recent findings},
author={Watkins, M. and Rumbelow, Jessica},
year={2023},
url={https://www.lesswrong.com/posts/Ya9LzwEbfaAMY8ABo/solidgoldmagikarp-ii-technical-details-and-more-recent},
publisher={lesswrong.com},
}

@misc{SolidGoldMagikarp3,
title={SolidGoldMagikarp III: Glitch token archaeology},
author={Watkins, M. and Rumbelow, Jessica},
year={2023},
url={https://www.lesswrong.com/posts/8viQEp8KBg2QSW4Yc/solidgoldmagikarp-iii-glitch-token-archaeology},
publisher={lesswrong.com},
}

@misc{Midjourney,
    title = "{Midjourney Documentation and User Guide}",
    url = "https://https://docs.midjourney.com/",
author={{Midjourney}},
}

@misc{Laion,
author={{LAION.ai}},
year={n.d.},
    title = "{LAION. Large-scale Artificial Intelligence Open Network}",
    URL= {https://laion.ai},
}

@inproceedings{NEURIPS2022a1859deb,
 author = {Schuhmann, Christoph and Beaumont, Romain and Vencu, Richard and Gordon, Cade and Wightman, Ross and Cherti, Mehdi and Coombes, Theo and Katta, Aarush and Mullis, Clayton and Wortsman, Mitchell and Schramowski, Patrick and Kundurthy, Srivatsa and Crowson, Katherine and Schmidt, Ludwig and Kaczmarczyk, Robert and Jitsev, Jenia},
 booktitle = {Advances in Neural Information Processing Systems},
 editor = {S. Koyejo and S. Mohamed and A. Agarwal and D. Belgrave and K. Cho and A. Oh},
 pages = {25278--25294},
 publisher = {Curran Associates, Inc.},
 title = {LAION-5B: An open large-scale dataset for training next generation image-text models},
 url = {https://proceedings.neurips.cc/paper_files/paper/2022/file/a1859debfb3b59d094f3504d5ebb6c25-Paper-Datasets_and_Benchmarks.pdf},
 volume = {35},
 year = {2022}
}

@inproceedings{feng-etal-2023-uncovering,
    title = "Uncovering Limitations in Text-to-Image Generation: A Contrastive Approach with Structured Semantic Alignment",
    author = "Feng, Qianyu  and
      Sui, Yulei  and
      Zhang, Hongyu",
    editor = "Bouamor, Houda  and
      Pino, Juan  and
      Bali, Kalika",
    booktitle = "Findings of the Association for Computational Linguistics: EMNLP 2023",
    month = dec,
    year = "2023",
    address = "Singapore",
    publisher = "Association for Computational Linguistics",
    url = "https://aclanthology.org/2023.findings-emnlp.595/",
    doi = "10.18653/v1/2023.findings-emnlp.595",
    pages = "8876--8888"
}

@article{KJMethod,
 ISSN = {00187259, 19383525},
 URL = {http://www.jstor.org/stable/44126786},
 author = {Raymond Scupin},
 journal = {Human Organization},
 number = {2},
 pages = {233--237},
 publisher = {Society for Applied Anthropology},
 title = {The KJ Method: A Technique for Analyzing Data Derived from Japanese Ethnology},
 urldate = {2025-05-10},
 volume = {56},
 year = {1997},
}

@misc{daras2022,
      title={Discovering the Hidden Vocabulary of DALLE-2}, 
      author={Giannis Daras and Alexandros G. Dimakis},
      year={2022},
      eprint={2206.00169},
      archivePrefix={arXiv},
}

@inproceedings{VIT,
title={An Image is Worth 16x16 Words: Transformers for Image Recognition at Scale},
author={Alexey Dosovitskiy and Lucas Beyer and Alexander Kolesnikov and Dirk Weissenborn and Xiaohua Zhai and Thomas Unterthiner and Mostafa Dehghani and Matthias Minderer and Georg Heigold and Sylvain Gelly and Jakob Uszkoreit and Neil Houlsby},
booktitle={International Conference on Learning Representations},
year={2021},
url={https://openreview.net/forum?id=YicbFdNTTy}
}

@InProceedings{HDBSCAN,
author="Campello, Ricardo J. G. B.
and Moulavi, Davoud
and Sander, Joerg",
editor="Pei, Jian
and Tseng, Vincent S.
and Cao, Longbing
and Motoda, Hiroshi
and Xu, Guandong",
title="Density-Based Clustering Based on Hierarchical Density Estimates",
booktitle="Advances in Knowledge Discovery and Data Mining",
year="2013",
publisher="Springer Berlin Heidelberg",
address="Berlin, Heidelberg",
pages="160--172",
isbn="978-3-642-37456-2"
}

@InProceedings{Rombach_2022_CVPR,
    author    = {Rombach, Robin and Blattmann, Andreas and Lorenz, Dominik and Esser, Patrick and Ommer, Bj\"orn},
    title     = {High-Resolution Image Synthesis With Latent Diffusion Models},
    booktitle = {Proceedings of the IEEE/CVF Conference on Computer Vision and Pattern Recognition (CVPR)},
    month     = {June},
    year      = {2022},
    pages     = {10684-10695}
}

@misc{janus,
      title={Janus-Pro: Unified Multimodal Understanding and Generation with Data and Model Scaling}, 
      author={Xiaokang Chen and Zhiyu Wu and Xingchao Liu and Zizheng Pan and Wen Liu and Zhenda Xie and Xingkai Yu and Chong Ruan},
      year={2025},
      eprint={2501.17811},
      archivePrefix={arXiv},
}

@misc{labs2025flux1kontextflowmatching,
      title={FLUX.1 Kontext: Flow Matching for In-Context Image Generation and Editing in Latent Space}, 
      author={Black Forest Labs and Stephen Batifol and Andreas Blattmann and Frederic Boesel and Saksham Consul and Cyril Diagne and Tim Dockhorn and Jack English and Zion English and Patrick Esser and Sumith Kulal and Kyle Lacey and Yam Levi and Cheng Li and Dominik Lorenz and Jonas Müller and Dustin Podell and Robin Rombach and Harry Saini and Axel Sauer and Luke Smith},
      year={2025},
      eprint={2506.15742},
      archivePrefix={arXiv},
      url={https://arxiv.org/abs/2506.15742}, 
}

@inproceedings{2205.11487.pdf,
title={Photorealistic Text-to-Image Diffusion Models with Deep Language Understanding},
author={Chitwan Saharia and William Chan and Saurabh Saxena and Lala Li and Jay Whang and Emily Denton and Seyed Kamyar Seyed Ghasemipour and Raphael Gontijo-Lopes and Burcu Karagol Ayan and Tim Salimans and Jonathan Ho and David J. Fleet and Mohammad Norouzi},
booktitle={Advances in Neural Information Processing Systems},
editor={Alice H. Oh and Alekh Agarwal and Danielle Belgrave and Kyunghyun Cho},
year={2022},
url={https://openreview.net/forum?id=08Yk-n5l2Al}
}

@InProceedings{CLIP,
  title = 	 {Learning Transferable Visual Models From Natural Language Supervision},
  author =       {Radford, Alec and Kim, Jong Wook and Hallacy, Chris and Ramesh, Aditya and Goh, Gabriel and Agarwal, Sandhini and Sastry, Girish and Askell, Amanda and Mishkin, Pamela and Clark, Jack and Krueger, Gretchen and Sutskever, Ilya},
  booktitle = 	 {Proceedings of the 38th International Conference on Machine Learning},
  pages = 	 {8748--8763},
  year = 	 {2021},
  editor = 	 {Meila, Marina and Zhang, Tong},
  volume = 	 {139},
  series = 	 {Proceedings of Machine Learning Research},
  month = 	 {18--24 Jul},
  publisher =    {PMLR},
  url = 	 {https://proceedings.mlr.press/v139/radford21a.html}
}

@inproceedings{samuel2023generatingimagesrareconcepts,
author = {Samuel, Dvir and Ben-Ari, Rami and Raviv, Simon and Darshan, Nir and Chechik, Gal},
title = {Generating images of rare concepts using pre-trained diffusion models},
year = {2024},
isbn = {978-1-57735-887-9},
publisher = {AAAI Press},
doi = {10.1609/aaai.v38i5.28270},
booktitle = {Proceedings of the Thirty-Eighth AAAI Conference on Artificial Intelligence and Thirty-Sixth Conference on Innovative Applications of Artificial Intelligence and Fourteenth Symposium on Educational Advances in Artificial Intelligence},
articleno = {522},
numpages = {9},
series = {AAAI'24/IAAI'24/EAAI'24}
}

@misc{williams2024drawlunderstandingeffectsnonmainstream,
      title={DrawL: Understanding the Effects of Non-Mainstream Dialects in Prompted Image Generation}, 
      author={Joshua N. Williams and Molly FitzMorris and Osman Aka and Sarah Laszlo},
      year={2024},
      eprint={2405.05382},
      archivePrefix={arXiv},
}

@inproceedings{2311.04287,
 author = {Lee, Tony and Yasunaga, Michihiro and Meng, Chenlin and Mai, Yifan and Park, Joon Sung and Gupta, Agrim and Zhang, Yunzhi and Narayanan, Deepak and Teufel, Hannah and Bellagente, Marco and Kang, Minguk and Park, Taesung and Leskovec, Jure and Zhu, Jun-Yan and Li, Fei-Fei and Wu, Jiajun and Ermon, Stefano and Liang, Percy S},
 booktitle = {Advances in Neural Information Processing Systems},
 editor = {A. Oh and T. Naumann and A. Globerson and K. Saenko and M. Hardt and S. Levine},
 pages = {69981--70011},
 publisher = {Curran Associates, Inc.},
 title = {Holistic Evaluation of Text-to-Image Models},
 url = {https://proceedings.neurips.cc/paper_files/paper/2023/file/dd83eada2c3c74db3c7fe1c087513756-Paper-Datasets_and_Benchmarks.pdf},
 volume = {36},
 year = {2023}
}

@inproceedings{3613904.3642803,
author = {Wang, Zhijie and Huang, Yuheng and Song, Da and Ma, Lei and Zhang, Tianyi},
title = {PromptCharm: Text-to-Image Generation through Multi-modal Prompting and Refinement},
year = {2024},
isbn = {9798400703300},
publisher = {Association for Computing Machinery},
address = {New York, NY, USA},
doi = {10.1145/3613904.3642803},
booktitle = {Proceedings of the 2024 CHI Conference on Human Factors in Computing Systems},
articleno = {185},
numpages = {21},
series = {CHI '24}
}

@misc{2408.12910,
      title={What Do You Want? User-centric Prompt Generation for Text-to-image Synthesis via Multi-turn Guidance}, 
      author={Yilun Liu and Minggui He and Feiyu Yao and Yuhe Ji and Shimin Tao and Jingzhou Du and Duan Li and Jian Gao and Li Zhang and Hao Yang and Boxing Chen and Osamu Yoshie},
      year={2024},
      eprint={2408.12910},
      archivePrefix={arXiv},
}

@misc{otani2023verifiablereproduciblehumanevaluation,
      title={Toward Verifiable and Reproducible Human Evaluation for Text-to-Image Generation}, 
      author={Mayu Otani and Riku Togashi and Yu Sawai and Ryosuke Ishigami and Yuta Nakashima and Esa Rahtu and Janne Heikkilä and Shin'ichi Satoh},
      year={2023},
      eprint={2304.01816},
      archivePrefix={arXiv},
}

@inproceedings{xu2023imagerewardlearningevaluatinghuman,
title={ImageReward: Learning and Evaluating Human Preferences for Text-to-Image Generation},
author={Jiazheng Xu and Xiao Liu and Yuchen Wu and Yuxuan Tong and Qinkai Li and Ming Ding and Jie Tang and Yuxiao Dong},
booktitle={Thirty-seventh Conference on Neural Information Processing Systems},
year={2023},
url={https://openreview.net/forum?id=JVzeOYEx6d}
}

@inproceedings{NEURIPS202373aacd8b,
 author = {Kirstain, Yuval and Polyak, Adam and Singer, Uriel and Matiana, Shahbuland and Penna, Joe and Levy, Omer},
 booktitle = {Advances in Neural Information Processing Systems},
 editor = {A. Oh and T. Naumann and A. Globerson and K. Saenko and M. Hardt and S. Levine},
 pages = {36652--36663},
 publisher = {Curran Associates, Inc.},
 title = {Pick-a-Pic: An Open Dataset of User Preferences for Text-to-Image Generation},
 url = {https://proceedings.neurips.cc/paper_files/paper/2023/file/73aacd8b3b05b4b503d58310b523553c-Paper-Conference.pdf},
 volume = {36},
 year = {2023}
}

@ARTICLE{2403.11821,
  author={Hartwig, Sebastian and Engel, Dominik and Sick, Leon and Kniesel, Hannah and Payer, Tristan and Poonam, Poonam and Glöckler, Michael and Bäuerle, Alex and Ropinski, Timo},
  journal={IEEE Transactions on Visualization and Computer Graphics}, 
  title={A Survey on Quality Metrics for Text-to-Image Generation}, 
  year={2025},
  volume={31},
  number={10},
  pages={9464-9483},
  doi={10.1109/TVCG.2025.3585077},
}

@article{Ringvold, title={AI Text-to-Image Generation in Art and Design Teacher Education: A Creative Tool or a Hindrance to Future Creativity?}, volume={1}, url={https://openjournals.ljmu.ac.uk/PATT40/article/view/1350}, number={October}, journal={The 40th International Pupils’ Attitudes Towards Technology Conference Proceedings 2023}, author={Ringvold, Tore Andre and Strand, Ingri and Haakonsen, Peter and Saasen Strand, Kari}, year={2023}, month={Oct.} }

@book{sperber1995relevance,
  title     = {Relevance: Communication and Cognition},
  author    = {Sperber, Dan and Wilson, Deirdre},
  year      = {1995},
  edition   = {2nd},
  publisher = {Blackwell},
  address   = {Oxford, UK},
}

@incollection{wilson2004relevance,
  title     = {Relevance Theory},
  author    = {Wilson, Deirdre and Sperber, Dan},
  booktitle = {The Handbook of Pragmatics},
  editor    = {Horn, Laurence R. and Ward, Gregory},
  pages     = {607--632},
  publisher = {Blackwell},
  address   = {Oxford, UK},
  year      = {2006},
doi={10.1002/9780470756959},
}

@ARTICLE{Vasilev2024S137,
author={Viacheslav Vasilev and Vladimir Arkhipkin and Julia Agafonova and Tatiana Nikulina and Evelina Mironova and Alisa Shichanina and Nikolai Gerasimenko and Mikhail Shoytov and Denis Dimitrov},
	title = {CRAFT: Cultural Russian-Oriented Dataset Adaptation for Focused Text-to-Image Generation},
	year = {2024},
	journal = {Doklady Mathematics},
	volume = {110},
	number = {Suppl 1},
	pages = {S137 – S150},
	doi = {10.1134/S1064562424602324},
}

@ARTICLE{Elsharif2025122636,
	author = {Elsharif, Wala and Alzubaidi, Mahmood and Agus, Marco},
	title = {Cultural Bias in Text-to-Image Models: A Systematic Review of Bias Identification, Evaluation, and Mitigation Strategies},
	year = {2025},
	journal = {IEEE Access},
	volume = {13},
	pages = {122636 – 122659},
	doi = {10.1109/ACCESS.2025.3585745},
}

@inproceedings{jha2024visageglobalscaleanalysisvisual,
    title = "{V}i{SAG}e: A Global-Scale Analysis of Visual Stereotypes in Text-to-Image Generation",
    author = "Jha, Akshita  and
      Prabhakaran, Vinodkumar  and
      Denton, Remi  and
      Laszlo, Sarah  and
      Dave, Shachi  and
      Qadri, Rida  and
      Reddy, Chandan K.  and
      Dev, Sunipa",
    editor = "Ku, Lun-Wei  and
      Martins, Andre  and
      Srikumar, Vivek",
    booktitle = "Proceedings of the 62nd Annual Meeting of the Association for Computational Linguistics (Volume 1: Long Papers)",
    month = aug,
    year = "2024",
    address = "Bangkok, Thailand",
    publisher = "Association for Computational Linguistics",
    url = "https://aclanthology.org/2024.acl-long.667/",
    doi = "10.18653/v1/2024.acl-long.667",
    pages = "12333--12347",
}

@misc{luccioni2023stablebiasanalyzingsocietal,
      title={Stable Bias: Analyzing Societal Representations in Diffusion Models}, 
      author={Alexandra Sasha Luccioni and Christopher Akiki and Margaret Mitchell and Yacine Jernite},
      year={2023},
      eprint={2303.11408},
      archivePrefix={arXiv},
}

@ARTICLE{vice2023quantifyingbiastexttoimagegenerative,
  author={Vice, Jordan and Akhtar, Naveed and Hartley, Richard and Mian, Ajmal},
  journal={IEEE Transactions on Dependable and Secure Computing}, 
  title={Quantifying Bias in Text-to-Image Generative Models}, 
  year={2025},
  volume={22},
  number={5},
  pages={5658-5671},
  doi={10.1109/TDSC.2025.3572115},
}

@inproceedings{10.1145/3706599.3719678,
author = {Lan, Xingyu and An, Jiaxi and Guo, Yisu and Chiyou, Tong and Cai, Xintong and Zhang, Jun},
title = {Imagining the Far East: Exploring Perceived Biases in AI-Generated Images of East Asian Women},
year = {2025},
isbn = {9798400713958},
publisher = {Association for Computing Machinery},
address = {New York, NY, USA},
url = {https://doi.org/10.1145/3706599.3719678},
doi = {10.1145/3706599.3719678},
booktitle = {Proceedings of the Extended Abstracts of the CHI Conference on Human Factors in Computing Systems},
articleno = {332},
numpages = {7},
series = {CHI EA '25}
}

@CONFERENCE{Kandwal202473,
	author = {Kandwal, Siddharth and Nehra, Vibha},
	title = {A Survey of Text-to-Image Diffusion Models in Generative AI},
	year = {2024},
	journal = {Proceedings of the 14th International Conference on Cloud Computing, Data Science and Engineering, Confluence 2024},
	pages = {73 – 78},
	doi = {10.1109/Confluence60223.2024.10463372},
}

@CONFERENCE{Saravanan202384,
	author = {Saravanan, Adhithya and Kocielnik, Rafal and Jiang, Roy and Han, Pengrui and Anandkumar, Anima},
	title = {Exploring Social Bias in Downstream Applications of Text-to-Image Foundation Models},
	year = {2023},
	journal = {Proceedings of Machine Learning Research},
	volume = {239},
	pages = {84 – 102},
}

@INPROCEEDINGS{11084610,
  author={Kuhn, Christopher B. and Ünver, Boğaç and Vera, Diego and Burgmair, Christoph and Aykut, Tamay},
  booktitle={2025 IEEE International Conference on Image Processing (ICIP)}, 
  title={Detecting and Mitigating Incoherent Input of Latent Diffusion Models}, 
  year={2025},
  volume={},
  number={},
  pages={337-342},
  doi={10.1109/ICIP55913.2025.11084610}}

@inproceedings{russo-2022-creative,
    title = "Creative Text-to-Image Generation: Suggestions for a Benchmark",
    author = "Russo, Irene",
    editor = {H{\"a}m{\"a}l{\"a}inen, Mika  and
      Alnajjar, Khalid  and
      Partanen, Niko  and
      Rueter, Jack},
    booktitle = "Proceedings of the 2nd International Workshop on Natural Language Processing for Digital Humanities",
    month = nov,
    year = "2022",
    address = "Taipei, Taiwan",
    publisher = "Association for Computational Linguistics",
    url = "https://aclanthology.org/2022.nlp4dh-1.18/",
    doi = "10.18653/v1/2022.nlp4dh-1.18",
    pages = "145--154",
}

@misc{LessWrong,
year={n.d.},
author={{LessWrong}},
    title = "{Glitch Tokens}",
publisher={LessWrong},
    url = "https://www.lesswrong.com/tag/glitch-tokens",
}

@misc{2507.17922.pdf,
      title={From Seed to Harvest: Augmenting Human Creativity with AI for Red-teaming Text-to-Image Models}, 
      author={Jessica Quaye and Charvi Rastogi and Alicia Parrish and Oana Inel and Minsuk Kahng and Lora Aroyo and Vijay Janapa Reddi},
      year={2025},
      eprint={2507.17922},
      archivePrefix={arXiv},
}

@article{10.1016/j.ijhcs.2024.103375,
author = {Rapp, Amon and Di Lodovico, Chiara and Torrielli, Federico and Di Caro, Luigi},
title = {How do People Experience the Images created by Generative Artificial Intelligence? An Exploration of People's Perceptions, Appraisals, and Emotions related to a Gen-AI Text-to-Image Model and its Creations},
year = {2025},
issue_date = {Jan 2025},
publisher = {Academic Press, Inc.},
address = {USA},
volume = {193},
number = {C},
issn = {1071-5819},
doi = {10.1016/j.ijhcs.2024.103375},
journal = {Int. J. Hum.-Comput. Stud.},
numpages = {16}
}

@inproceedings{Liu2022,
    author = {Liu, Vivian and Chilton, Lydia B.},
    title = {Design Guidelines for Prompt Engineering Text-to-Image Generative Models},
    year = {2022},
    isbn = {9781450391573},
    publisher = {Association for Computing Machinery},
    address = {New York, NY, USA},
    doi = {10.1145/3491102.3501825},
    booktitle = {Proceedings of the 2022 CHI Conference on Human Factors in Computing Systems},
    numpages = {23}
}

@article{Trope2010-TROCTO,
	author = {Yaacov Trope and Nira Liberman},
	doi = {10.1037/a0018963},
	journal = {Psychological Review},
	number = {2},
	pages = {440--463},
	title = {Construal-Level Theory of Psychological Distance},
	volume = {117},
	year = {2010}
}

@inproceedings{ko2023large,
author = {Ko, Hyung-Kwon and Park, Gwanmo and Jeon, Hyeon and Jo, Jaemin and Kim, Juho and Seo, Jinwook},
title = {Large-scale Text-to-Image Generation Models for Visual Artists' Creative Works},
year = {2023},
isbn = {9781450399685},
publisher = {Association for Computing Machinery},
address = {New York, NY, USA},
doi = {10.1145/3581641.3584078},
booktitle = {Proceedings of the 28th International Conference on Intelligent User Interfaces},
pages = {919–933},
numpages = {15},
series = {IUI '23}
}

@article{reber2004processing,
author = {Reber, Rolf and Schwarz, Norbert and Winkielman, Piotr},
title = {Processing Fluency and Aesthetic Pleasure: Is Beauty in the Perceiver's Processing Experience?},
journal = {Personality and Social Psychology Review},
volume = {8},
number = {4},
pages = {364--382},
year = {2004},
doi = {10.1207/s15327957pspr0804_3},
publisher = {SAGE Publications}
}

@inproceedings{kurosu1995apparent,
author = {Kurosu, Masaaki and Kashimura, Kaori},
title = {Apparent Usability vs. Inherent Usability: Experimental Analysis on the Determinants of the Apparent Usability},
year = {1995},
isbn = {0897917553},
publisher = {Association for Computing Machinery},
address = {New York, NY, USA},
doi = {10.1145/223355.223680},
booktitle = {Conference Companion on Human Factors in Computing Systems},
pages = {292–293},
numpages = {2},
series = {CHI '95}
}

@incollection{wertheimer1923gestalt,
  author = {Wertheimer, Max},
  title = {Laws of Organization in Perceptual Forms},
  booktitle = {A Source Book of Gestalt Psychology},
  editor = {Ellis, W. D.},
  pages = {71--88},
  year = {1938},
  publisher = {Routledge and Kegan Paul},
  address = {London},
  note = {English translation of Wertheimer, M. (1923). Untersuchungen zur Lehre von der Gestalt II. Psychologische Forschung, 4, 301--350.}
}

@book{bartlett1932remembering,
  author = {Bartlett, Frederic C.},
  title = {Remembering: A Study in Experimental and Social Psychology},
  year = {1995},
  publisher = {Cambridge University Press},
  address = {Cambridge, UK},
}

\appendix
\section*{Appendix}
\section{Prompts}
\label{appendix:tokens}
{%
\noindent
A1 -- Unique Names:
    \textit{Aranoak Iarnin,
    Arend Perlaft,
    Aziel Vartani,
    Droak Fedril,
    Fizzah Borandur,
    Frentl Geddik,
    Fraoa Seinalur,
    Jhureth Nyxar,
    Kaithis Vornex,
    Meztli Acan,
    Nuoka Zuberi,
    Oarn Iznilaker,
    Papatya Marwood,
    Pirnak Zerlian,
    Shelan Creswell,
    Soraya Lonescu,
    Torrence Vexley,
    Tzivani Rhekai,
    Unnah Nenosim,
    Uua Lakeiram}
}%

{%
\vspace{.3\baselineskip}
\noindent
A2 -- Corrupted Words:
    \textit{Greagoft,
    Greafirx,
    Greawarz,
    Greatird,
    Greanist,
    Grealorn,
    Greamylt,
    Greavorn,
    Greaweld,
    Greatrix,
    Shuttastfr,
    Shuttagpdr,
    Shuttaaips,
    Shuttagust,
    Shuttamagu,
    Shuttagait,
    Shuttapmt,
    Shuttagolt,
    Shuttaiain,
    Shuttapirn}
}%

{%
\vspace{.3\baselineskip}
\noindent
A3 -- URLs:
    \textit{\nolinkurl{www.amazon.com},
    \nolinkurl{www.bing.com},
    \nolinkurl{www.chatgpt.com},
    \nolinkurl{www.deepseek.com},
    \nolinkurl{www.ebay.com},
    \nolinkurl{www.facebook.com},
    \nolinkurl{www.imdb.com},
    \nolinkurl{www.instagram.com},
    \nolinkurl{www.linkedin.com},
    \nolinkurl{www.nytimes.com},
    \nolinkurl{www.paypal.com},
    \nolinkurl{www.pinterest.com},
    \nolinkurl{www.tiktok.com},
    \nolinkurl{www.tumblr.com},
    \nolinkurl{www.quora.com},
    \nolinkurl{www.reddit.com},
    \nolinkurl{www.whatsapp.com},
    \nolinkurl{www.wikipedia.org},
    \nolinkurl{www.x.com},
    \nolinkurl{www.youtube.com}
    }%
}%

{%
\vspace{.3\baselineskip}
\noindent
A4.1 -- Finnish:
\textit{    mökki,
    tuoli,
    pöytä,
    kasvi,
    kauppa,
    maito,
    kinkku,
    muovi,
    mustikka,
    suola,
    hevonen,
    heinasirkka,
    lapanen,
    saapas,
    pelto,
    lato,
    käsi,
    iho,
    akseli,
    pultti}
}%

{%
\vspace{.3\baselineskip}
\noindent
A4.2 -- Tagalog:
\textit{%
apoy,
aso,
bata,
daga,
dila,
galit,
gising,
gubat,
iyak,
kamay,
kulay,
lakad,
lupa,
matanda,
ngiti,
paa,
pagpag,
takbo,
tulog,
ulan    
    }%
}%

{%
\vspace{.3\baselineskip}
\noindent
A5 -- Glitch Tokens:
\textit{    Rawdownload,
    PsyNetMessage,
    UnfocusedRange,
    Downloadha,
    RandomRedditorWithNo,
    InstoreAndOnline,
    externalActionCode,
    Embedreportprint,
    ExternalToEVAOnly,
    cloneembedreportprint
    }
}%

{%
\vspace{.3\baselineskip}
\noindent
A6 -- Abbreviations:
\textit{    bcd,
    dwew,
    fsde,
    gsod,
    sfew,
    gogw,
    pqwe,
    fsie,
    dqae,
    gafc,
    grd,
    jfdsk,
    tkrafl,
    fkeriw,
    ltr,
    dfaf,
    fdfd,
    rtyu,
    fgetiu,
    daqz}
}%

\section{Dataset}
\label{appendix:dataset}

\begin{table}[!bth]
\caption{Description of dataset columns}
\label{tab:datasetcolumns}
\small
\begin{tabularx}{\linewidth}{lXX}
\toprule
    Column & Description & Example \\
\midrule
    Content     & Raw message collected from Midjourney, including the prompt, parameters, and username & **painted rose petals --v 4** - @voltz23 (fast) \\
    Prompt      & Pre-processed prompt, extracted from Content & painted rose petals \\
    User        & Discord username, extracted from Content & voltz23 \\
    Image       & Path to the image extracted from the four-image panel & dataset/images/general-19/voltz23\_painted\_rose\_petals\_4a1b15f9-ecf5--9CD4A\_2.png \\
    Num         & Number of image (1=top left, 2=top right, 3=bottom left, 4=bottom right) & 2 \\
    Reactions   & Reactions to the message from Discord users & 
    $\ast$ (1) \\
    Version     & The Midjourney version, detected in the prompt or inferred by date & 4.0 \\ 
    Date        & Date and time of message & 2023-02-01T17:53:24.1370000+00:00 \\
    Datetime    & Date and time without timezone & 2023-02-01 17:53:24.137 \\
    Lang        & Detected prompt language & en \\
    Channel     & Name of Midjourney discord channel & general-19 \\
\bottomrule
\end{tabularx}
\end{table}

\section{Additional Examples of Default Images}%
\label{appendix:more-examples}%
%
\begin{figure*}[!h]
  \centering
    \includegraphics[trim=1.6mm 0 0 0,clip,width=.8\textwidth]{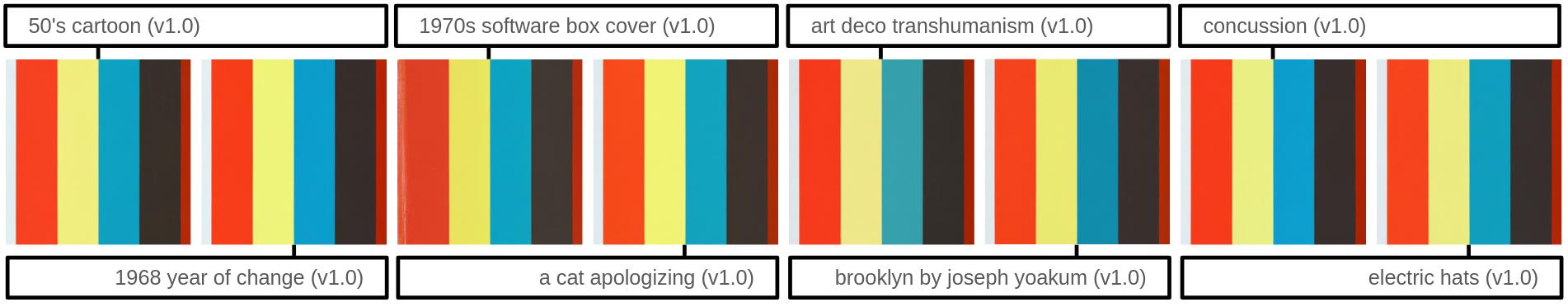}%
\\[.15\baselineskip]%
    \includegraphics[trim=1.6mm 0 0 0,clip,width=.8\textwidth]{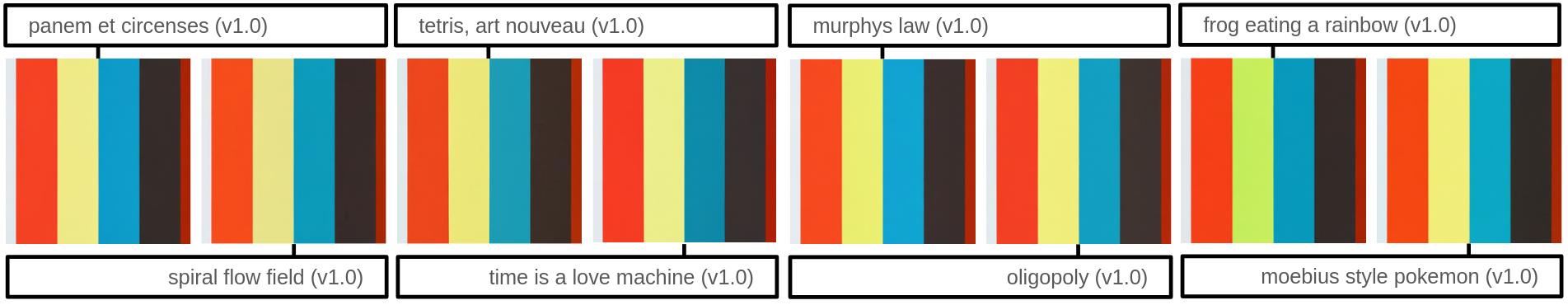}%
\\[.15\baselineskip]%
    \includegraphics[width=.8\textwidth]{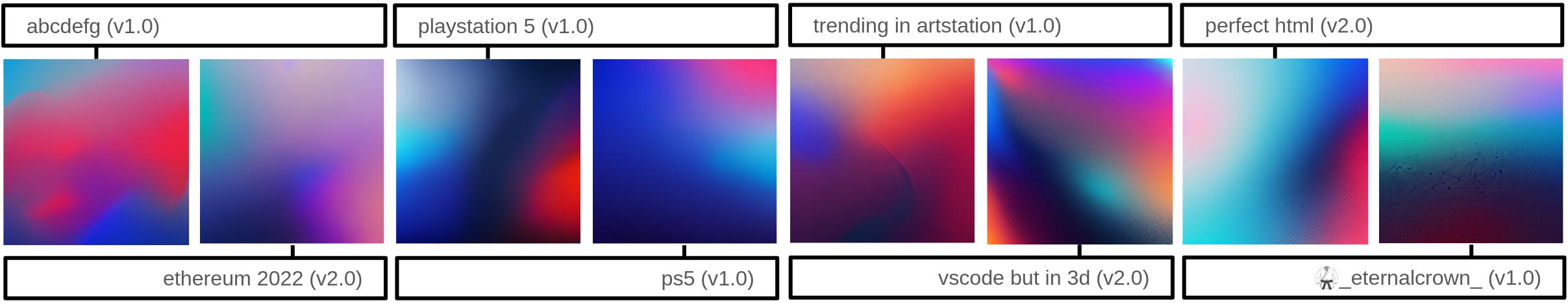}%
\\[.15\baselineskip]%
    \includegraphics[width=.8\textwidth]{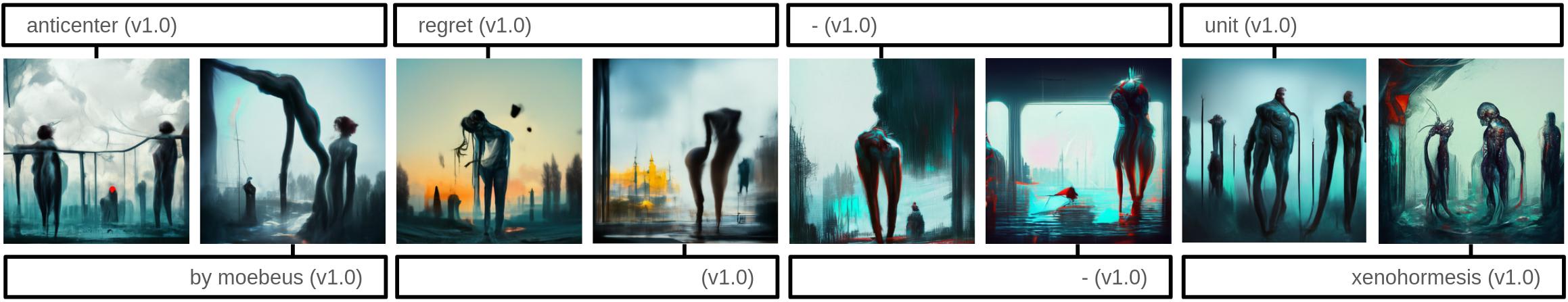}%
\\[.15\baselineskip]%
    \includegraphics[width=.8\textwidth]{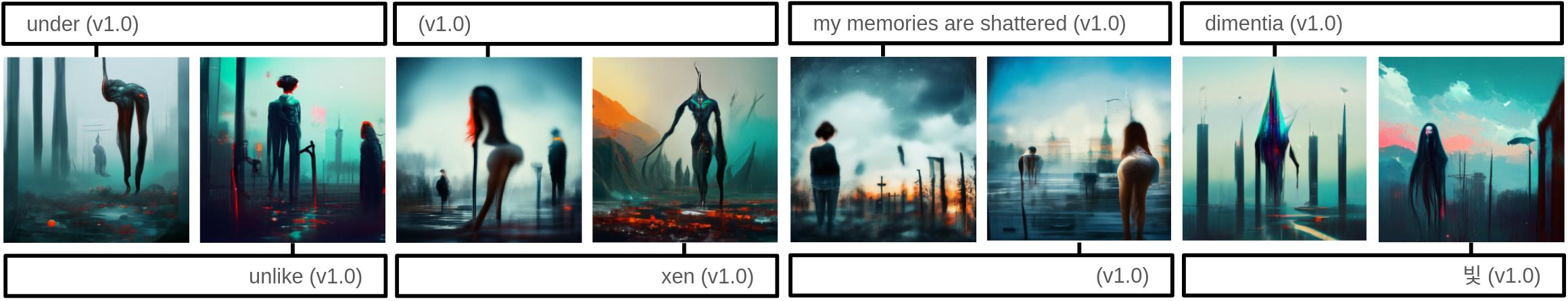}%
\\[.15\baselineskip]%
    \includegraphics[width=.8\textwidth]{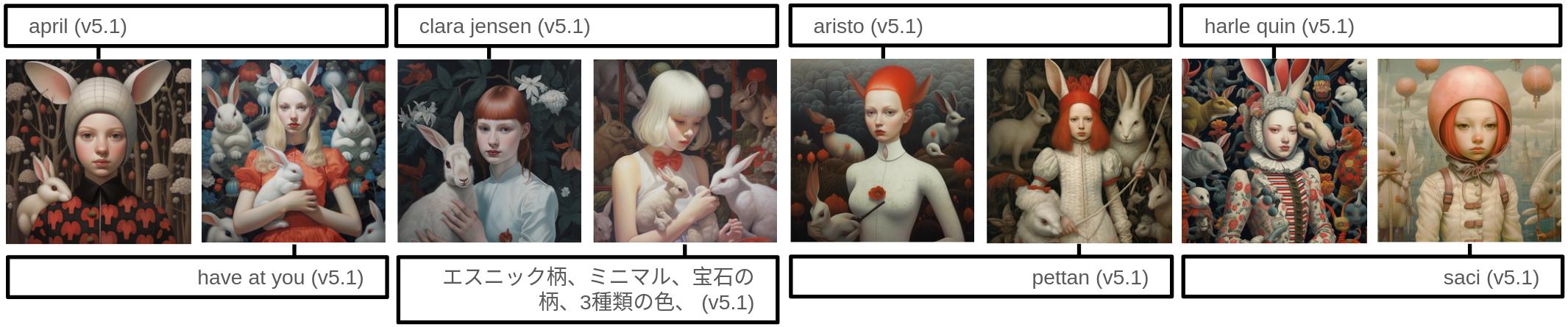}%
\\[.15\baselineskip]%
\caption{Additional examples of default images}%
\Description{The figure depicts sets of default images from different Midjourney versions. The visual similarity in terms of subject, colors, style, and composition is uncanny.}%
\label{fig:more-results}%
\end{figure*}%

\begin{figure*}[!h]\ContinuedFloat
  \centering
  \captionsetup{list=off}
    \includegraphics[width=.8\textwidth]{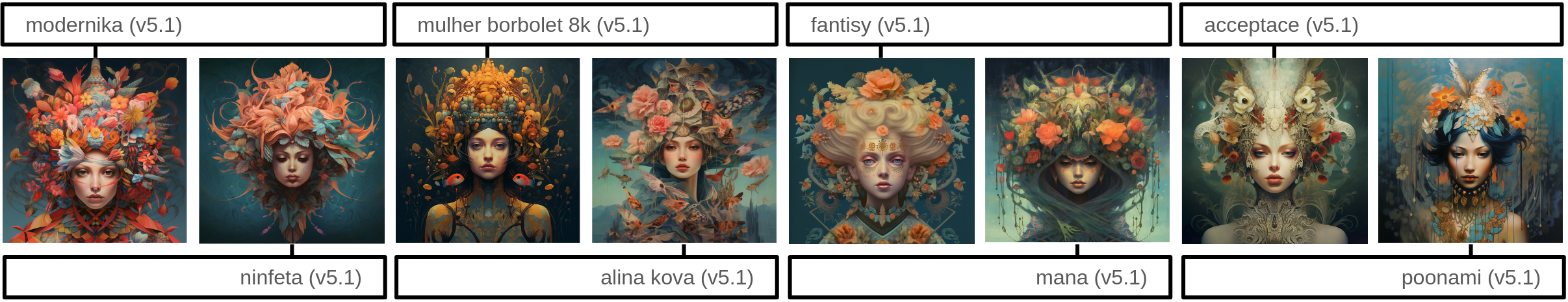}%
\\[.15\baselineskip]%
    \includegraphics[width=.8\textwidth]{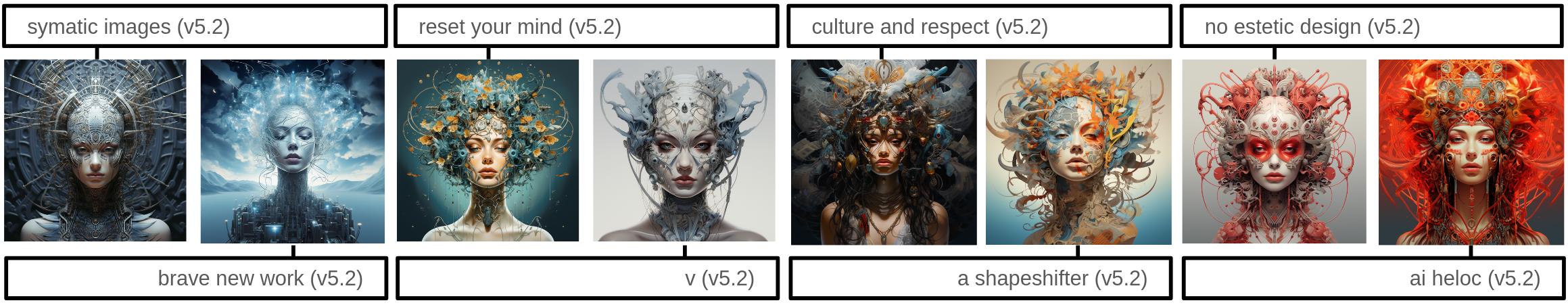}%
\\[.15\baselineskip]%
    \includegraphics[width=.8\textwidth]{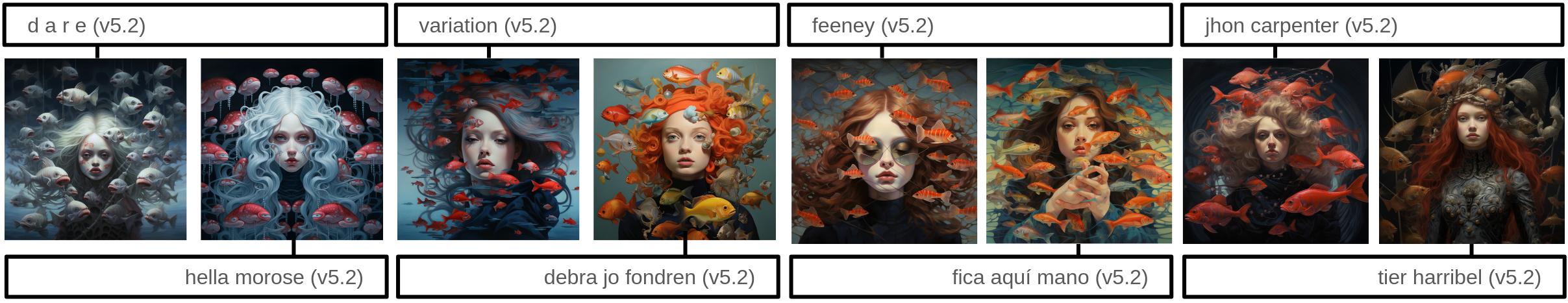}%
\\[.15\baselineskip]%
    \includegraphics[width=.8\textwidth]{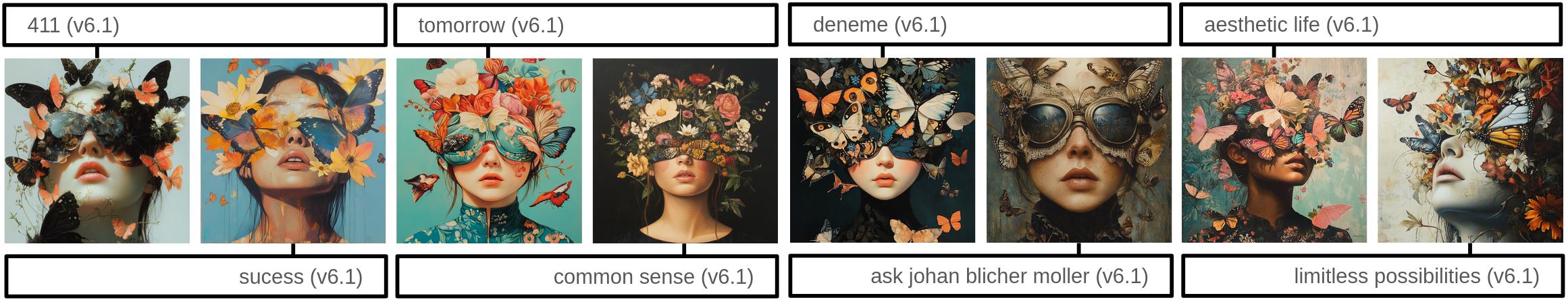}%
\\[.15\baselineskip]%
    \includegraphics[width=.8\textwidth]{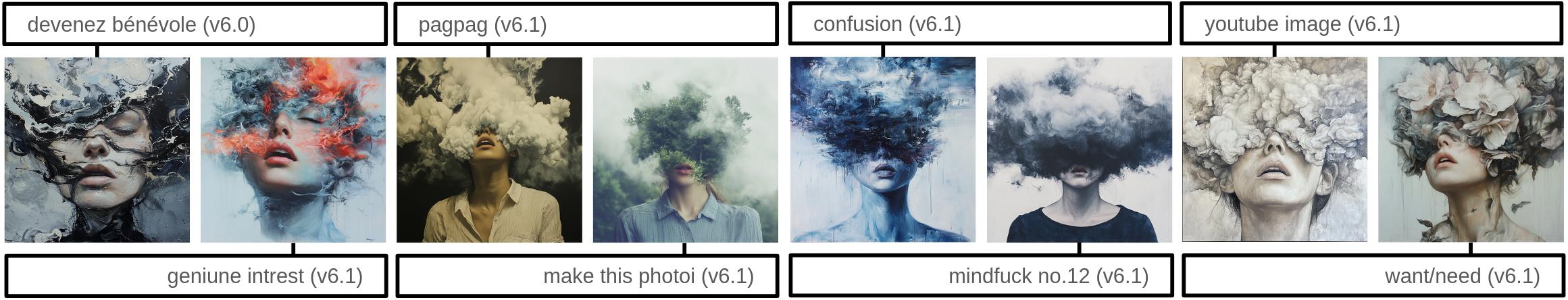}%
\caption{Additional examples of default images (continued)}%
\end{figure*}

\end{document}